\documentclass{WileyMSP-template}
\usepackage{amsmath}
\usepackage{amsfonts}
\usepackage{amssymb}
\usepackage{graphicx}
\usepackage{mathrsfs}
\usepackage{braket}
\usepackage{dcolumn} 
\usepackage{bm}
\usepackage{epstopdf}
\usepackage{color}
\usepackage{xcolor}
\usepackage{lipsum}
\usepackage{afterpage}
\usepackage{ragged2e}
\usepackage{tabularx}
\begin{document}

\pagestyle{fancy}
\rhead{\includegraphics[width=2.5cm]{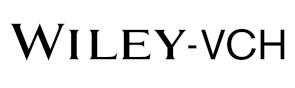}}

\title{A hybrid graphene-siliconnitride nanomembrane as a versatile and ultra-widely tunable mechanical device}

\maketitle

\author{Mengqi Fu,}
\author{Bojan Bo{\v{s}}njak,}
\author{Zhan Shi,}
\author{Jannik Dornseiff,}
\author{Robert H. Blick,}
\author{Elke Scheer, and}
\author{Fan Yang*}


\dedication{~}

\begin{affiliations}
Dr. Mengqi Fu, MSc Jannik Dornseiff, Prof. Dr. Elke Scheer\\
Fachbereich Physik, Universit{\"a}t Konstanz, 78457 Konstanz, Germany\\

MSc Bojan Bosnjak, Prof. Dr. Robert H. Blick\\
Center for Hybrid Nanostructures, Universit{\"a}t Hamburg, 22761 Hamburg, Germany\\

Dr. Zhan Shi\\
Department of Mechanics, Key Laboratory of Soft Machines and Smart devices of Zhejiang Province, Zhejiang University, 310058 Hangzhou, China

Dr. Fan Yang\\
Fachbereich Physik, Universit{\"a}t Konstanz, 78457 Konstanz, Germany\\
Dynamic Precision Micro\&Nano Sensing Technology Research Institute, Chongqing, 400030 Chongqing, China\\
Email Address: fan.yang@uni-konstanz.de\\
\end{affiliations}

\keywords{graphene nanomembrane, thermal-electro-mechanics, metal-graphene-siliconnitride hybrid structure, mechanical properties regulation, frequency tuning, symmetry breaking, controlled motion and actuation}

\begin{abstract}

Integration of 2D materials in nanoelectromechanical systems (NEMS) marries the robustness of silicon-based materials with exceptional electrical controllability in 2D materials, drastically enhancing system performance which now is the key for many advanced applications in nanotechnology. Here, we experimentally demonstrate and theoretically analyze a powerful on-chip graphene integrated NEMS device consisting of a hybrid graphene/silicon-nitride membrane with metallic leads that enables an extremely large static and dynamic parameter regulation. When a static voltage is applied to the leads, the force induced by the thermal expansion difference between the leads and the membrane results in ultra-wide frequency tuning, deformation (post-buckling transition) and regulation of mechanical properties. Moreover, by injecting an alternating voltage to the leads, we can excite the resonator vibrating even far beyond its linear regime without a complex and space consuming actuation system. Our results prove that the device is a compact integrated system possessing mechanical robustness, high controllability, and fast response. It not only expands the limit of the application range of NEMS devices but also pushes multidimensional nanomechanical resonators into working in the nonlinear regime.\\

\end{abstract}

\section{Introduction}
Nanoelectromechanical systems (NEMS) play an essential role in sensing, actuation, and filtering applications \cite{cleland1998nanometre, halg2021membrane, lemme2020nanoelectromechanical, samanta2022low, tong2022advanced, zou2022aluminum}, in fundamental researches such as optomechanics \cite{thompson2008strong, barzanjeh2022optomechanics, brawley2016nonlinear}, nonlinear dynamics \cite{bachtold2022mesoscopic, yang2019spatial, yang2021mechanically, ochs2022frequency, guttinger2017energy}, 2D materials \cite{vsivskins2020magnetic, steeneken2021dynamics} and other fields of physics \cite{shin2022spiderweb, de2016tunable, karg2020light}. Regulating the mechanical properties, e.g. the eigenfrequency and spatial symmetry, over a large range is central in NEMS for their application, e.g. for temperature drift compensation \cite{wang2020temperature}, adjusting their application range, and for controlling their nonlinear behavior \cite{eichler2013symmetry, ochs2021resonant, samanta2018tuning, kozinsky2006tuning, piller2020thermal}. Moreover, the programmable regulation of spatial symmetry enables application as nano-actuators \cite{majstrzyk2018thermomechanically, samanta2022low} for morphology changes applied in flexible devices \cite{yang2019flexible} and nanorobotics \cite{huang2016soft}. However, typical on-chip regulating structures, such as a capacitive gates or comb structures have limited tuning ability and require complex design and fabrication techniques.\\
\indent NEMS structures that utilize the electrothermomechanic (ETM) effect \cite{manoach2004coupled,xu2019piezoresistive, pruchnik2022study} have been shown to be more effective for tuning the eigenfrequency \cite{zhang2021thermal, chiout2023extreme, steeneken2021dynamics}, the nonlinearity \cite{suo2022tuning}, and for system actuation capabilities \cite{samanta2022low, safouene2013voltage, xu2019piezoresistive, majstrzyk2018thermomechanically}. An example is the Timoshenko beam, i.e. a bilayered structure composed of different materials \cite{timoshenko1925analysis} that is the building block of standard thermally activated switches.
NEMS resonators built from two-dimensional (2D) materials show large tuning effects \cite{chiout2023extreme, OpticalcontrolMoS2, ye2021ultrawide}, but are fragile and have a lower quality factor $Q$ than quasi-2D Si-based resonators that instead feature robust elastic properties but less tuning capability \cite{sementilli2022nanomechanical, yang2023quantitative}. Therefore, an appropriate assembly of these two concepts in combination with the ETM effect can improve the regulation function and may result in robust high-performance nanomechanical devices.\\
\indent In this work, we present a metal-graphene-siliconnitride (MGS) hybrid structure combined with an ETM tuning scheme to mechanically deform a freestanding siliconnitride (SiN) membrane and control the eigenfrequency and the nonlinearity of the resonator over an extremely wide range. Furthermore, we demonstrate that the MGS device features on-surface vibration excitation with the capability of controlled symmetry breaking. We discuss the ETM actuation mechanism and quantitatively demonstrate the excellent performance of the symmetry breaking control and on-surface actuation of the MGS resonator.\\

\section{Results and discussion}
\subsection{Design concept of the MGS device}
\indent The designed MGS resonator, as depicted in \textbf{Fig. \ref{fig:Setup}} (a), is based on a silicon chip that carries a fully-clamped freestanding SiN membrane. To induce ETM effects locally on the membrane, a graphene-metal (G-M) structure is placed on the SiN membrane. The patterned monolayer G is located centrally on the SiN membrane, serving as the conductive channel. The M (Ti/Au, 25/25 nm thick) leads are evaporated on top for providing electrical contacts and for ETM actuation. The cross section of the MGS resonator is presented in Fig. \ref{fig:Setup} (b). \\
\indent The designed ETM actuation scheme is based on the 
force between the M leads and the SiN membrane induced by their different thermal expansion coefficients. Hence we choose a material combination that features a large gradient of thermal expansion coefficients from the top Au to the bottom SiN ($\alpha_\textrm{Au} = 14.2 \cdot 10^{-6}$/K, $\alpha_\textrm{Ti} = 7.6 \cdot 10^{-6}$/K, and $\alpha_\textrm{SiN} = 3.2 \cdot 10^{-6}$/K). Since the thermal expansion itself will be caused by Joule heating, we take advantage of the high electrical conductance of the G and the relatively high contact resistance of the M-G interface to localize the Joule heating to the small overlap area between  G and the M leads.
The local voltage drop and related dissipation at the contacts causes the ETM effect that tunes the mechanical properties of the device, as shown in Fig.~\ref{fig:Setup} (b). It can both statically deflect the SiN membrane by applying a direct voltage $V_{\textrm{DC}}$, as shown in Fig.~\ref{fig:Setup} (c), and dynamically excite the vibration of the membrane by applying an alternating voltage $V_{\textrm{AC}}$, as shown in Fig.~\ref{fig:Setup} (d). \\
\indent In the device under study here, shown in Fig.~\ref{fig:Setup} (e), featuring 5 leads, labeled 1 to 5, the resistance of the G is about 1.65\,k$\Omega$ and the contact resistance $R$ is about 5 to 10 times larger, varying from lead to lead and the applied voltage. The SiN membrane has a lateral size of 495\,$\times$\,512\,$\mu$m$^2$ and a thickness of $\sim$ 110\,nm, fabricated using a backside wet-etching procedure. The G is square-shaped with $300 \times 300 \mu$m$^2$. For more details, see the Supporting Information (SI) \cite{SI}.\\
\indent Additionally, the high flexibility, small bending stiffness, and ultra-low mass of the G sheet are beneficial for the mechanics. Since the bending stiffness and mass of the G are negligible compared to the ones of the SiN membrane \cite{waitz2012mode,zhang2015vibrational}, it provides a quasi-free end for the M leads on the surface. The mechanical properties of this NEMS device are dominated by the SiN membrane and the ETM effects are regulated by the properties of the MGS structure. In addition, the high optical transparency of G ensures that the optical properties of the SiN are minimally affected, which is beneficial for the application in fields such as optomechanics.\\
\subsection{Static spatial symmetry control}
\begin{figure}[thp]
  \centering
  \includegraphics[width=0.6\linewidth]{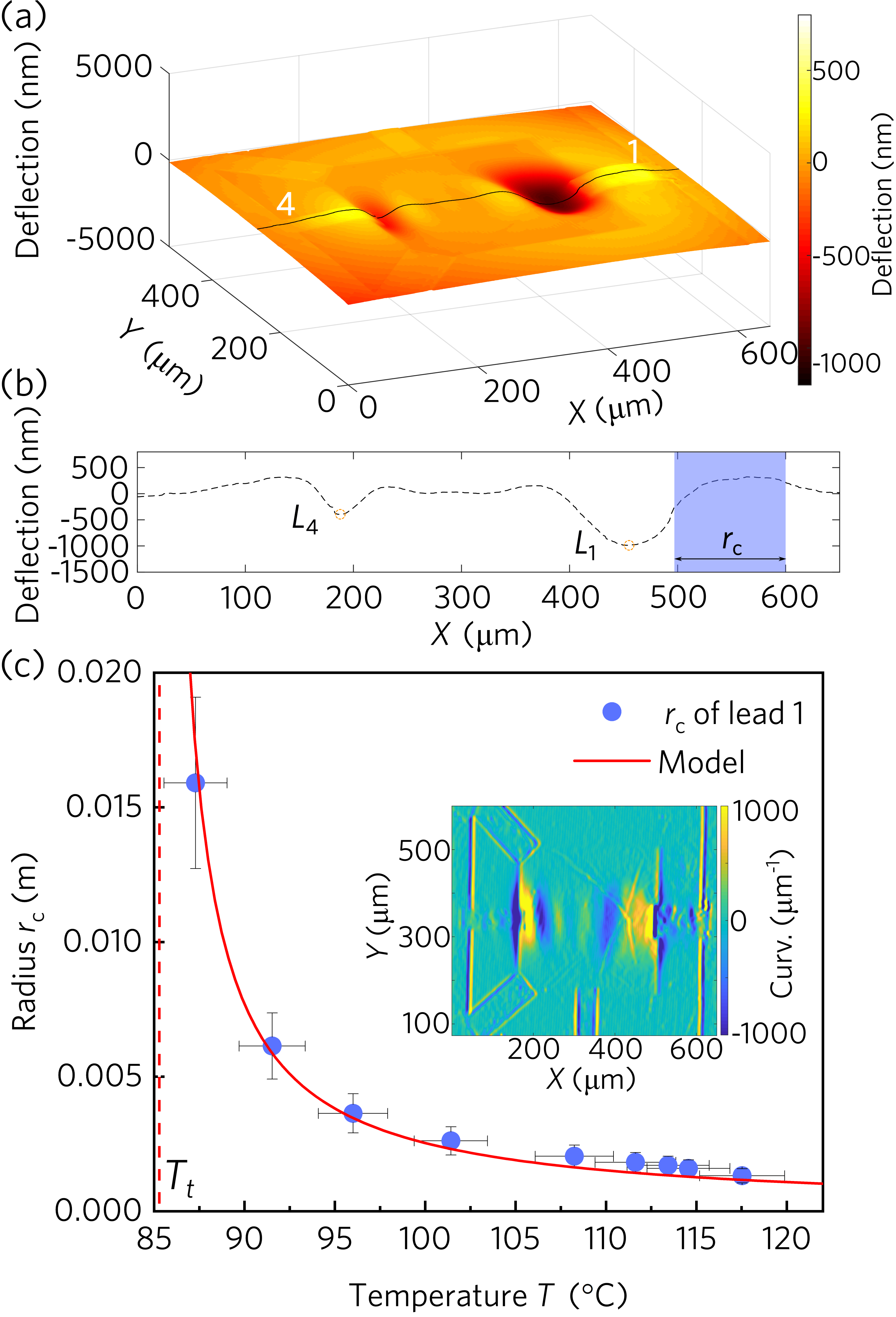}
  \caption{Experimental measurement of controlled static deformation and extracted curvature radius $r_c$ of the deflection. 
  (a) IWLI-captured image of the spatial deflection of the MGS resonator with $V_\textrm{DC}$ = 2.9\,V. (b) Deflection profile of the resonator at the position on the membrane surface indicated by the black line in (a). The two maxima of the deflected SiN membrane marked as $L_1$ and $L_4$ correspond to the positions of the open ends of leads 1 and 4, respectively. The spatial range used for calculating  $r_c$ of the MGS structure is indicated by the blue-shaded area. (c) Extracted $r_c$ plotted as function of the corresponding temperature (heated up by $V_\textrm{DC}$ from 2.65\,V up to 2.9\,V, blue dots), the calculation of the bilayer model is plotted as red solid line. The red-dashed line showing the buckling transition temperature $T_t$. The inset shows the curvature distribution extracted from the IWLI image obtained for $V_\textrm{DC}=2.9$\,V.
  }\label{fig:strain_radii}
\end{figure}

%
%
\indent We use leads 1 and 4 for applying the $V_\textrm{DC}$. Up to approximately 2.5\,V, the MGS structure remains flat. An out-of-plane static deflection of the MGS device occurs when $V_\textrm{DC}$ exceeds approximately 2.5\,V, corresponding to the scheme in Fig.~\ref{fig:Setup} (e). The amplitude and the curvature of the static deflection increase proportionally to $V_\textrm{DC}$, as shown in the SI \cite{SI}. A typical 3D image of the static spatial deflection captured by imaging white light interferometry (IWLI) at $V_\textrm{DC}$ = 2.9\,V with a line cut parallel to the $x$-axis is shown in \textbf{Fig.~\ref{fig:strain_radii}} (a) and (b). The strongest downward deflections of the device are located around the free ends of leads 1 and 4. The MGS structures are convexly bent upwards, indicated by the bright yellow structures in Fig.~\ref{fig:strain_radii} (a). 
The maximum deflection, beyond 1300\,nm, occurs around the edge of lead 1 (point $L_1 \approx 450~\mu$m in the $x$-axis in Fig.~\ref{fig:strain_radii} (b)). The images under different $V_\textrm{DC}$ show similar bending patterns but different amplitudes of deflection, as shown in the SI \cite{SI}. When the $V_\textrm{DC}$ is switched off, the system recovers its original flat state, proving that the deflection is reversible over this range of $V_\textrm{DC}$.\\
\indent The very large spatial deflection of the MGS resonator under $V_\textrm{DC}$ can be fully described by the buckling transition induced by unequal thermal expansion of the M leads and the SiN membrane, as explained below \cite{xiao2024dual, ritchie1975buckling}.
When the MGS structure senses heat, the M layer elongates and generates a force component perpendicular to the SiN plane, henceforth referred to as buckling force, arising from the directly exerted linear force component. 
(sliding is prevented). After a certain threshold, the MGS structure spontaneously bends, acting as a typical Timoshenko bilayer structure \cite{timoshenko1925analysis, SI}. The M leads bend downwards at the free end, resulting in a detectable curvature and deformation of the SiN layer.\\
\indent We calculated the relationship between Joule heating and the curvature radius $r_c$ for a simplified model representing one side of the MGS resonator (e.g., lead 1) as a uniformly heated bilayer cantilever with a stress constantly applied on the free end. Since the Ti and Au layers have similar thickness, we use a simplified metal film model with averaged material parameters (thermal properties). In addition, We assume that the majority of the heat is generated at the interface between the M lead and the graphene and that the metal leads are heated evenly due to their high thermal conductivity, allowing us to estimate the temperature at different $V_\textrm{DC}$ with Fourier's law:
\begin{equation}
\label{Fourier_law}
T = T_0 + \frac{L_M P_M}{\kappa W_M h_M},
\end{equation}
where $T_0$ is the starting temperature (here: room temperature), $L_M$, $W_M$, and $h_M$ denote the length, width and thickness of the M lead acting as heat source. 
$\kappa$ is the thermal conductivity of the M lead, and $P_M$ is the power load to the structure with the M-G contact resistance $R$. Here, considering the resistance contribution of the G and the contact resistance, $P_M$ is calculated as 42.5\% of the total applied power $P = IV$. This value takes into account that in total about 85\% of the electrically dissipated power contributes to the heating of the membrane and that both leads contribute equally to the heating.\\

The curvature of the MGS at different temperature $T$ follows the general equation:
%
\begin{equation}
\label{eq_temp-curvature}
r_c = \frac{(h_M+h_{SiN})\left[3(1+h')^2+(1+h'E')\left(h'^2+\frac{1}{h'E'}\right)\right]} {(6\varepsilon \cdot (T-T_0)-\sigma_{SiN}/E_{SiN})(1+h')^2},
\end{equation}
where $h'=h_{M}/h_{SiN}$, $E' = E_{M}/E_{SiN}$, and $\varepsilon = \alpha_M - \alpha_{SiN}$. Here, $h_{SiN}$ and $h_M$ represent the thicknesses of SiN and Au/Ti, respectively. $E_{SiN}$, $E_M$, $\alpha_{SiN}$, and $\alpha_M$ represent the Young's moduli and thermal expansion coefficients of the two layers, respectively. $\sigma_{SiN}$ represents the residual stress of SiN. For more detailed information, we refer to the SI \cite{SI}.\\
The $r_c$ values extracted from the experiment as a function of temperature ($V_\textrm{DC}$ from 2.5\,V up to 2.9\,V, applied between leads 1 and 4 are shown as blue dots in Fig.~\ref{fig:strain_radii} (c), while the calculated $r_c$ using Eq. \eqref{eq_temp-curvature} is shown as a red solid line. The curvature distribution of the membrane at $V_\textrm{DC}$ = 2.9\,V is shown in the inset of Fig.~\ref{fig:strain_radii} (c) for a full visualization of the spatial deflection. The results demonstrate that the calculated $r_c$ matches the experimental results well without any free fitting parameter, supporting the notion that the strong mismatch of the thermal expansion coefficients of the different materials in the MGS structure is the origin of the static deformation, which can be engineered and predicted accurately. A finite element analysis also proves the consistency, for details see the SI \cite{SI}.\\
The model also provides access to the buckling transition temperature,  required to initiate the spatial deflection,  $T_t = (\sigma_2/6\varepsilon E_{SiN}) + T_0 \simeq 85~$\textcelsius $~$ ($V_\textrm{DC} \gtrsim$ 2.5\,V). $T_t$ is obtained from Eq.~\eqref{eq_temp-curvature} as the temperature at which the first factor in the denominator vanishes and $r_c$ hence diverges. At $T_t$, marked in Fig.~\ref{fig:strain_radii} (c), the ETM induced force is exactly compensated by the local residual stress. Below $T_t$, no spatial deflection is detectable.
More importantly, the model reveals how $T_t$ can be engineered by designing the material properties and geometry, including residual stress, Young's modulus, and the thermal expansion coefficient of each component of the hybrid structure.\\
\indent In general, creating such strong and controlled static deformation in SiN membranes is difficult to achieve due to their high prestress and stiffness, especially when using on-chip structures such as dielectric gates. Most impressively, the ETM-induced strain of the MGS device can be extremely large and even break the SiN membrane, as demonstrated in the SI \cite{SI}. \\
\subsection{Regulation of dynamic properties}
\begin{figure}[thp]
  \centering
  \includegraphics[width=0.55\linewidth]{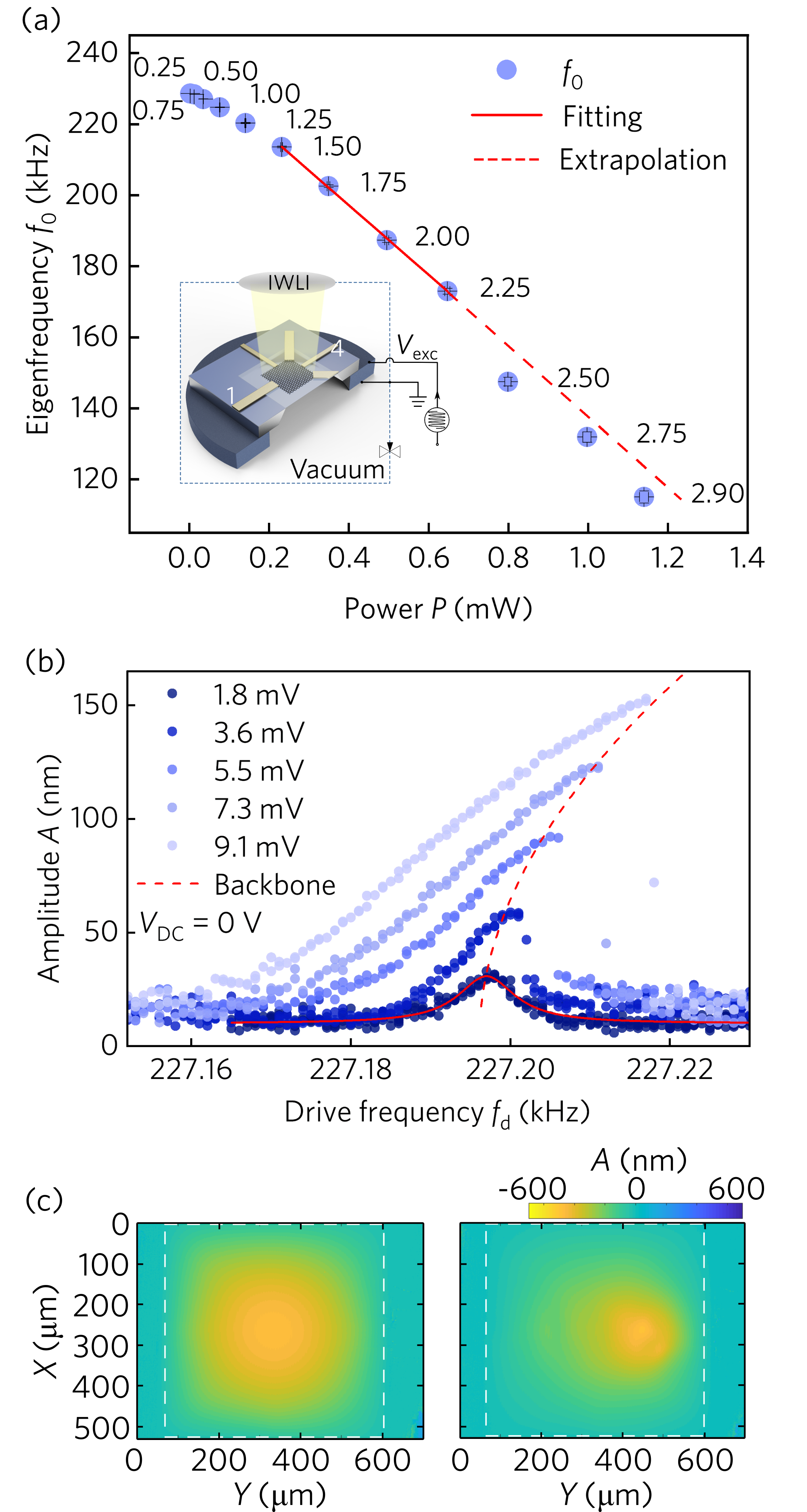}
  \caption{Experimental demonstration of dynamic properties tuning. (a) Tuning of the linear eigenfrequency $f_0$ of the (1,1) mode with increasing DC power $P$ loaded between leads 1 and 4. A linear fit is applied to the data with $0.2 \leq P \leq $ 0.64\,mW and plotted as a red-dashed line extended to the full data range. The inset illustrates the measurement scheme for obtaining the linear mechanical response of the MGS resonator. An AC sinusoidal drive signal $V_\textrm{exc}$ is sent into the piezo attached to the bottom of the chip to drive the resonator into vibration. Part of the structure is cut out to show the 3D spatial arrangement. (b) Nonlinear mechanical vibrational response curves of the MGS resonator with $V_\textrm{exc}$ applied to the piezo from 1.8\,mV up to 9.1\,mV with $V_\textrm{DC} = 0$\,V. (c) shows the IWLI-captured vibrational pattern of the (1,1) mode at phase $\phi = 90^{\circ}$ under $V_\textrm{DC}$ = 0\,V (left) and 2.75\,V (right), respectively. The $V_\textrm{DC}$ is applied to leads 1 and 4. The MGS device is driven by the piezo with $V_\textrm{exc} = 0.5$\,V and the $f_d$ is slightly above the respective $f_0$.}  
  \label{fig:RU_eigen_shift}
\end{figure}

\indent We now discuss the dynamic property tuning utilizing the ETM effect. The dynamic properties of the MGS device, such as the eigenfrequency $f_0$, the damping rate, the nonlinearity and the vibrational pattern, can be characterized from their vibration response. In our setup, we solidly attach the MGS resonator onto a piezo ring (serving as drive system), as shown in the inset of \textbf{Fig. \ref{fig:RU_eigen_shift}} (a). The mechanical response of the driven motion is measured optically by IWLI. The drive frequency $f_d$ at which the maximal vibration amplitude in the linear response is observed is identified as $f_0$. By fitting the linear response with a Lorentzian function and the nonlinear response with Duffing model, respectively, the damping rate (and thus the quality factor $Q$) as well as the cubic nonlinearity can be quantitatively extracted. For instance, Fig. \ref{fig:RU_eigen_shift} (b) shows a series of mechanical response curves ($V_\textrm{DC} = 0$\,V) for the (1,1) mode of the MGS resonator measured by IWLI. With $V_\textrm{exc}$ ranging from 1.8\,mV to 9.1\,mV, the response curves vary from a Lorentzian shape to a Duffing-type shape and the backbone trace is indicated by a red-dashed line. The $f_0$ and $Q$ are determined as 227.196 kHz and $2.8\times10^4$ from the response curve with $V_\textrm{exc}$ = 1.8\,mV in Fig. ~\ref{fig:RU_eigen_shift} (b)). The Duffing nonlinearity is extracted as $3.5\times10^{22} ~\textrm{m}^{-2}\textrm{s}^{-2}$ (see SI \cite{SI}).\\
\indent We first monitor the frequency tuning effect by characterizing the $f_0$ shift of the driven mode. Figure \ref{fig:RU_eigen_shift} (a) shows the dependence of the $f_0$ of the ground mode ((1,1) mode) on the power load $P$ with $V_\textrm{DC}$ applied between leads 1 and 4. $f_0$ decreases roughly linearly for increasing $P$ up to 0.64\,mW ($V_\textrm{DC}=$ 2.25\,V), as shown by the red dashed line. The damping rate remains almost unchanged, see the SI \cite{SI}. At $P \simeq 0.79$ mW ( $V_\textrm{DC} \simeq$ 2.50\,V), $f_0$ shows a  jump-like decrease and then follows the linear trend again with very similar slope as before the jump. This jump occurs at the power load of the onset of spatial deflection at $T = T_t \simeq 85$ \textcelsius. Upon increasing $P$ to 1.16\,mW (corresponding to $V_\textrm{DC}$ = 2.9\,V), $f_0$ decreases to 110 kHz, indicating a tuning of more than $50\%$. The offset in the linear tendency reveals the additional effect of the buckling on the geometric properties and intensifies the tuning of $f_0$ by $P$.\\
\indent Moreover, the vibrational pattern of the MGS resonator can be tuned notably by utilizing the ETM effect when buckling sets in. Fig. \ref{fig:RU_eigen_shift} (c) shows the comparison between the vibrational patterns of the (1,1) mode captured before ($V_\textrm{DC}$ = 0\,V, left panel) and after ($V_\textrm{DC}$ = 2.75\,V, right panel) the buckling transition for the phase $\phi=90^{\circ}$ recorded by stroboscopic illumination.
The spatial deflection of the vibrational motion under $V_\textrm{DC}$ = 0\,V (left panel) shows a usual sinusoidal envelope of the amplitude distribution, indicating a negligible asymmetry of the system. In contrast, the envelope of the amplitude distribution deviates from the sinusoidal shape at $V_\textrm{DC}$ = 2.75\,V, as shown in the right panel. The area with the maximal vibration amplitude shifts from the center to the area of lead 1, demonstrating a controlled symmetry breaking. This significant symmetry breaking can also be explained by the ETM effect: the heating above the buckling transition changes the internal stress distribution inhomogeneously resulting in an asymmetric stress distribution in the MGS device \cite{kozinsky2006tuning, ochs2021resonant}.\\
\subsection{On-surface actuation of a MGS resonator}
\begin{figure}[thp]
  \centering
  \includegraphics[width=0.55\linewidth]{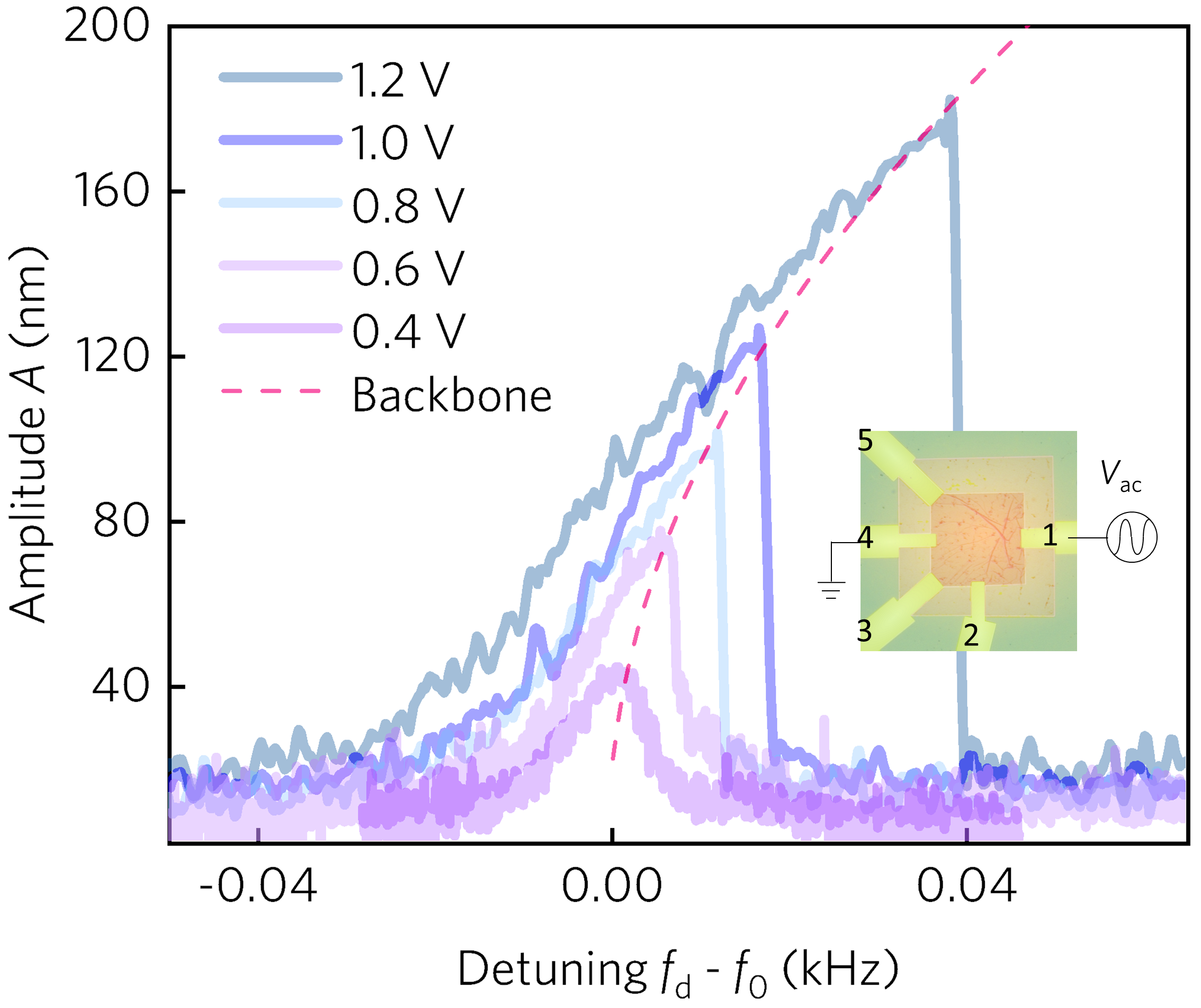}
  \caption{Nonlinear dynamic response curves of the MGS resonator actuated by the  ETM effects with different input voltages $V_\textrm{AC}$ ranging from 0.4\,V to 1.2\,V and measured by IWLI. The backbone trace is plotted as a red dashed line. The inset shows the circuit for the ETM drive, where no piezo is applied.} \label{fig:AC_strain_drive}
\end{figure}

\indent Furthermore, the ETM effect also provides an easy way to realize on chip driving of resonances, corresponding to the scheme in Fig. \ref{fig:Setup} (f). The resonator can be driven by applying an AC voltage $V_\textrm{AC}$ between leads 1 and 4, without any external (piezo) drive  as shown in the inset of \textbf{Fig. \ref{fig:AC_strain_drive}}, where a series of resonance curves ($V_\textrm{AC}$ from 0.4\,V to 1.2\,V ) is  plotted as a function of the detuning. 
At $V_\textrm{AC}$ = 0.4\,V, the $Q$ factor of the resonance is 2.3 $\times 10^4$, which is similar to the one obtained from the piezo-driven response in the linear regime at $V_\textrm{DC}$ = 0\,V in Fig. \ref{fig:RU_eigen_shift} (b).
Increasing $V_\textrm{AC}$ drives the system into the nonlinear regime, resulting in a slight shift of $f_0$ of the (1,1) mode under different $V_\textrm{AC}$. The Duffing nonlinearity of the resonator is determined to be $\gamma = 2.5 \times 10^{22} ~\textrm{m}^{-2}\textrm{s}^{-2}$ when $V_\textrm{AC}$ is smaller than 1.0\,V .\\
\indent When $V_\textrm{AC}$ increases to 1.2 V , the nonlinearity increases due to the ETM-induced strain of the resonator. 
The ETM-induced strain modifies the potential symmetry, leading to the appearance of an asymmetric term in the mode's potential energy \cite{ochs2021resonant}. This results in a cubic nonlinearity of the potential, i.e. a quadratic term in the equation of motion. The dynamics with broken symmetry at $V_\textrm{AC}$ = 1.2 V  can be well described by the potential: 
\begin{equation}
\label{eq:potential_sym_brk}
V(q) = \frac{2\pi f_{0}^2}{2}q^2 + \frac{\gamma_{3}}{3}q^3 + \frac{\gamma_{4}}{4}q^4 .
\end{equation}
Here, $q$ represents the deflection amplitude, and the potential includes a cubic nonlinearity ($\gamma_3$) and a quartic (Duffing) nonlinearity ($\gamma_{4}$). The parameter $\gamma_3$ is related to the symmetry breaking. By solving the corresponding equation of motion, we extracted $\gamma_3 = 9.65 \times 10^{17} ~\textrm{m}^{-1}\textrm{s}^{-2}$ and $\gamma_4 = 5.38 \times 10^{23} ~\textrm{m}^{-2}\textrm{s}^{-2}$. The value of $\gamma_4$ is one order of magnitude larger than the Duffing nonlinearity ($\gamma$) extracted previously, meaning that symmetry breaking significantly influences the nonlinearity. 
As a result, the presented MGS structure utilizing ETM effects provides new insights into controllable symmetry breaking through on-surface structures for silicon-based dielectric membrane resonators. Furthermore, the engineered tunable symmetry breaking enables the MGS structure to be highly versatile, thereby enabling numerous applications.\\
\indent Furthermore, we wish to comment on the robustness and versatility of our method for resonators. Previous methods for tuning the eigenfrequency, nonlinearity, vibrational pattern, and symmetry breaking require separate structures, such as a bottom and/or side gate \cite{kozinsky2006tuning, ochs2021resonant}, an attached piezo actuation system \cite{yang2021mechanically}, or separate laser illumination \cite{OpticalcontrolMoS2}. In contrast, our MGS structure provides an on-surface integration possibility with an extremely large capability for tuning and driving. The ETM method relies solely on the parameters $R$, $f_0$, $\Gamma$, $P$ ($V_\textrm{DC}$ or $V_\textrm{AC}$), geometry, and material properties. All of these parameters can be directly extracted from experimental characterization or from a one-time fitting of the driven modes, making our method much simpler and more straightforward. Additionally, the ETM method provides a direct way to control the dynamic behavior under different symmetry conditions. Finally, we note that the dynamic response reported here has been recorded at the same frequency as the drive frequency $f_d$ and not at its harmonic $2f_d$, as one could expect due the heat-related mechanism. We argue that differences of the M-G contacts result in a different heat dissipation for the two current directions which gives therefore rise to $1f$ periodicity of the excitation. We believe that there will be also a $2f$ component that, however, we cannot capture with our IWLI. \\
\section{Conclusion}
\indent In this study, we designed and fabricated a NEMS device composed of a MGS system bearing the potential for actuation and wide-range tuning. We clarify the corresponding mechanisms and models for the MGS device under different working scenarios. Firstly, by applying a DC voltage to the MGS structure, we can control the static deformation of the NEMS device by utilizing the ETM effect, and the corresponding bilayer model is well established. This provides a direct way of motion and actuation control, especially for 2D-NEMS devices. Secondly, in a NEMS resonator, the MGS structure achieves an extremely large tuning capability (over $50\%$) of the eigenfrequency. There are two main contributions: one from the generated heat, and the other from the ETM-induced static deformation. Thirdly, by applying an AC voltage, on-surface driving of the membrane resonator can be achieved without external excitation. The AC modulated ETM effect excites the vibration of the resonator, even into the nonlinear regime. Finally, also the nonlinear properties can be significantly tuned by the symmetry breaking, and are quantitatively described by the Duffing nonlinear model extended by an additional cubic nonlinearity of the potential. The developed MGS structure opens up possibilities for NEMS actuation applications and breaks the tuning limit of NEMS devices, also for advanced applications of 2D nonlinear resonators. Examples include controlled mode coupling/decoupling \cite{OpticalcontrolMoS2}, effective nonlinearity cancellation \cite{ochs2021resonant}, and energy conservation for the amplitude stabilization in persistent response without working in the nonlinear parametric coupling regime \cite{yang2021persistent}.\\


\section{Experimental Section}

\begin{figure}[thp]
  \centering
  \includegraphics[width=\textwidth]{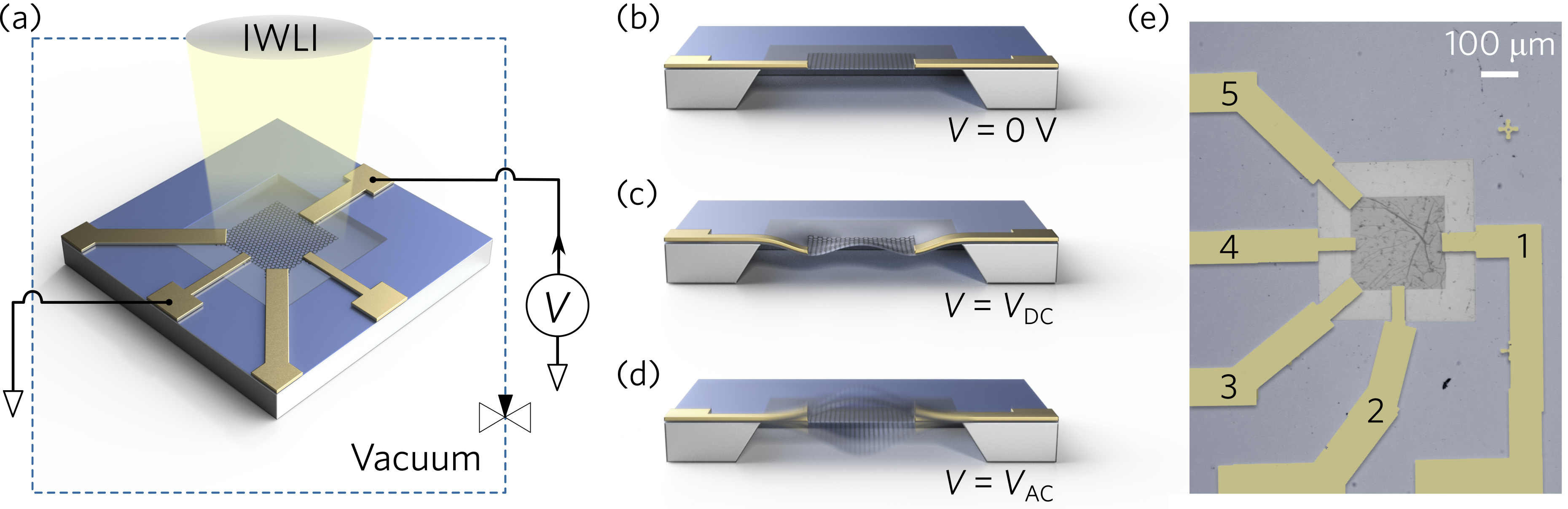}
  \caption{Experimental scheme and demonstration of the concept. (a) A 3D sketch of the MGS device, the electric circuit, and the IWLI. The graphene area is fixed by the leads, the SiN membrane holds the G and M structures and is clamped by a Si frame. (b)-(d) Simplified cross-section sketches of the MGS device (only one pair of leads: 1 and 4, see e)) for different conditions: b) plain surface for $V = 0$, c) static deformation with $V = V_\textrm{DC}$, and d) periodic vibration with $V = V_\textrm{AC}$. (e) Optical micrograph of the fabricated MGS device. The metal leads are labeled from 1 to 5.}
  \label{fig:Setup}
\end{figure}

\threesubsection{Sample fabrication}
\indent The SiN membrane is produced by wet-etching a 0.5\,mm thick commercial (110) silicon wafer coated with a $\mathrm{\sim}$110\,nm thick layer of LPCVD SiN on both sides in aqueous KOH. The sample presented here has a lateral size of 495 $\mu$m $\times$ 512 $\mu$m, and a thickness of 110\,nm. The monolayer G is grown by chemical vapor deposition and transferred to the surface of the SiN membrane using a wet method \cite{bosnjak2017investigating}, covering the entire SiN membrane as well as part of the Si frame. 
The G is then patterned by using electron beam lithography to define the channel and leads, followed by an oxygen plasma etching to remove the unwanted parts of G. Next, the lead areas of G are contacted with Ti/Au top electrodes (total thickness of about 50\,nm, marked from 1 to 5 in Fig.~\ref{fig:Setup} (e)) patterned by electron beam lithography and deposited by electron beam evaporation.)\\
\threesubsection{Measurement scheme}
The chip carrying the SiN sample is fixed to a piezo ring. The sample is placed in a vacuum chamber at room temperature with a base pressure $p \simeq 1\times 10^{-2}$\,mbar. 
\\
\indent The surface of the membrane is characterized using IWLI with various light sources, as described in detail in \cite{petitgrand2013d}. Dynamic deflections of the membrane can be excited by applying an AC voltage $V_\textrm{exc} \cdot \sin (2\pi f_dt$) to the piezo, resulting in an inertial excitation of the membrane. The excitation voltage is applied using a sinusoidal function generator, the phase of which can be synchronized with the stroboscopic light of the interferometer. In the study of the dynamic properties of MGS structure continuous light was used to record the resonance curves \cite{yang2023quantitative}, while stroboscopic light was used to measure the vibrational patterns \cite{waitz2012mode}. 
Examples of different mode shapes are presented in the SI \cite{SI}. The mechanical properties (Young's modulus and residual stress) are determined from the dispersion relation as described in \cite{waitz2012mode, waitz2015spatially}.\\
The two-point current-voltage ($I-V$) curves are slightly nonlinear, which suggests that the contact resistances are not negligible. The two-point resistance of the G-M structure decreases from several tens of k$\Omega$ to around 10\,k$\Omega$ as $V_\textrm{DC}$ is increased from 0 to 3\,V, depending on the lead combination, see the SI \cite{SI}. The resistance of the G channel is determined by the van-der-Pauw measurement \cite{ramadan1994van} and amounts to $R= 1.65\,\textrm{k}\Omega$ for the present sample, see the SI \cite{SI}.
\\
Further details about the sample fabrication, setup, and fitting processes can be found in our previous works \cite{yang2023quantitative, waitz2012mode, waitz2015spatially}.

\medskip
\textbf{Supporting Information} \par 
Supporting Information is available from the Wiley Online Library or from the authors.

\medskip
\textbf{Acknowledgements} \par 
The authors thank M. Dykman, E. Weig, W. Belzig, V. Gusev, O. Zilberberg, Y. Jiang, D. Bone\ss, H. Gu, Y. Jiang and Y. Peng for fruitful discussion and comments about the work. The authors gratefully acknowledge financial support from the Deutsche Forschungsgemeinschaft  (DFG, German Research Foundation) through Project-ID 425217212 (SFB 1432) and project 510766045, and the Wisconsin Alumni Research Foundation (WARF) via the Accelerator Program.\\
\medskip

%
%
%
\bibliographystyle{MSP}
\bibliography{Dispersion}
\newpage

%
%
%
%
\clearpage
%
%
%
\clearpage
%
%
\clearpage
%
%
%
%
%
%
\begin{figure}
\textbf{Table of Contents}\\
\medskip
  \includegraphics{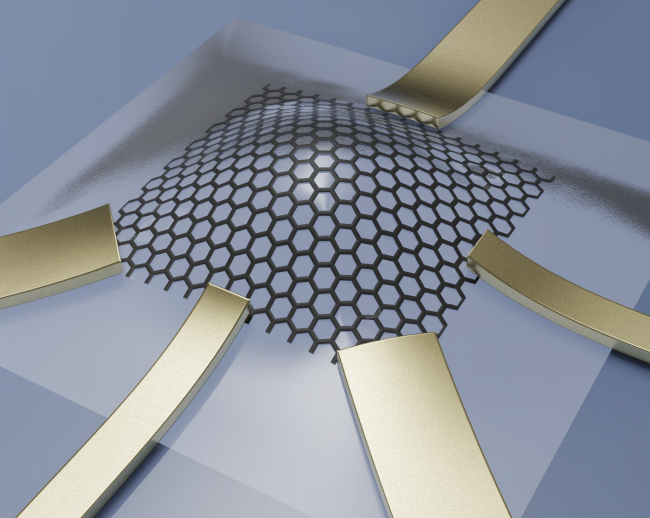}
  \medskip
  \caption*{The graphene integrated NEMS device consists of a hybrid graphene/silicon-nitride
membrane with metallic leads that enables ultra-wide frequency tuning, spatial deflection, mechanical properties tuning and on-surface actuation.}
\end{figure}

\end{document}


\title{Supporting Information of: A hybrid graphene-siliconnitride nanomembrane as a versatile and ultra-widely tunable mechanical device}
\author{Mengqi Fu}
\affiliation{Fachbereich Physik, Universit{\"a}t Konstanz, 78457 Konstanz, Germany}
\author{Bojan Bo{\v{s}}njak}%
\affiliation{Center for Hybrid Nanostructures, Universit{\"a}t Hamburg, 22761 Hamburg, Germany}
\author{Zhan Shi}%
\affiliation{Department of Mechanics, Key Laboratory of Soft Machines and Smart devices of Zhejiang Province, Zhejiang University, 310058 Hangzhou, China}
\author{Jannik Dornseiff}
\affiliation{Fachbereich Physik, Universit{\"a}t Konstanz, 78457 Konstanz, Germany}
\author{Robert H. Blick}%
\affiliation{Center for Hybrid Nanostructures, Universit{\"a}t Hamburg, 22761 Hamburg, Germany}
\author{Elke Scheer}%
\affiliation{Fachbereich Physik, Universit{\"a}t Konstanz, 78457 Konstanz, Germany}
\author{Fan Yang}
 \email{fan.yang@uni-konstanz.de}
\affiliation{Fachbereich Physik, Universit{\"a}t Konstanz, 78457 Konstanz, Germany}
\affiliation{Dynamic Precision Micro\&Nano Sensing Technology Research Institute, Chongqing, 400030 Chongqing, China}
\maketitle
%
%
%
%
%
%
\subsection{Sample fabrication and setup}
\subsubsection{Sample fabrication}
\noindent  The SiN membrane is produced by wet-etching a 0.5\,mm thick commercial (100) silicon wafer coated with a $\mathrm{\sim}$ 110 nm thick layer of LPCVD SiN on both sides in aqueous KOH. The sample presented here has a lateral size of 495\,$\mu$m $\times$ 512$~\mu$m, and a thickness of 110\,nm. The monolayer G is grown by chemical vapor deposition and transferred to the surface of the SiN membrane using a wet method \cite{bosnjak2017investigating}, covering the entire SiN membrane as well as part of the Si frame after the wet transfer. 

\begin{figure}[thbp]
   \includegraphics[width=0.4\linewidth]{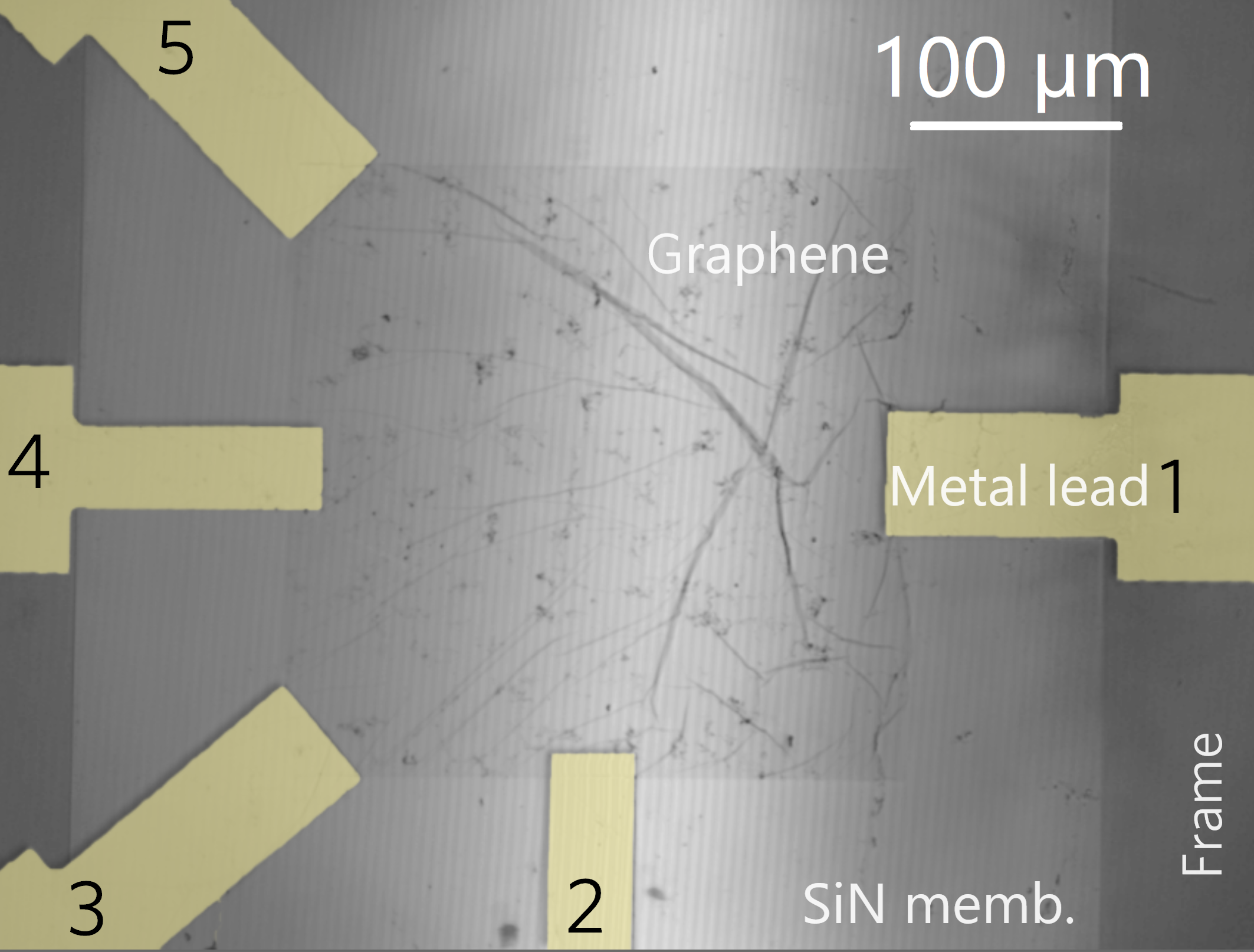}
   \caption[width=0.5\textwidth]
   {Image of the MGS device captured by IWLI. The metal leads are marked from 1 to 5.}
   \label{fig:RU_IWLI}
 \end{figure}

To pattern the G and thus define the G channel as well as the contact area, a layer of negative resist is spin-coated on the surface of G and patterned by the electron beam lithography. An oxygen plasma etching process is then performed to etch away the unwanted parts of G. After removing the negative resist in hot acetone, the metal lead area is patterned by a standard electron beam lithography process using a positive-resist system (one MMA-MAA copolymer layer and one PMMA ayer). A layer of Ti/Au (thicknesses around 25 nm each) metal film is evaporated by electron beam evaporation followed by a lift-off process. The metal leads are marked from 1 to 5 in Fig.~\ref{fig:RU_IWLI} as well as in the Fig.~1(e) in the main text. 
Further details about the sample fabrication, setup, and fitting processes can be found in our previous works \cite{yang2023quantitative, waitz2012mode, waitz2015spatially}.\\
\subsubsection{Measurement scheme}

\noindent The sample with SiN membrane and G devices is glued to a printed circuit board (PCB) which has connections to external electronics. The metal leads are bonded to the PCB, so that an electric signal can be applied to them. The PCB is glued to a piezo ring that can vibrate perpendicular to the surface of the Si substrate of the sample under an AC voltage. Note that this piezo ring is only used in the measurements of the eigenfrequency and the nonlinearity of the SiN membrane under $V_\textrm{DC}$ in this work. The sample is placed in a vacuum chamber at room temperature. The chamber is connected to a pump which keeps the $p$ = 0.001 mbar for the experiments. For $p \leq 0.01$ mbar, the damping is dominated by intrinsic damping mechanisms of the membrane and its clamping, and losses due to coupling to the atmosphere are negligible \cite{yang2023quantitative}. 
The surface of the membrane is examined using IWLI with various light sources, as described in detail in \cite{petitgrand2013d}. 

\begin{figure}[thbp]
   \includegraphics[width=\linewidth]{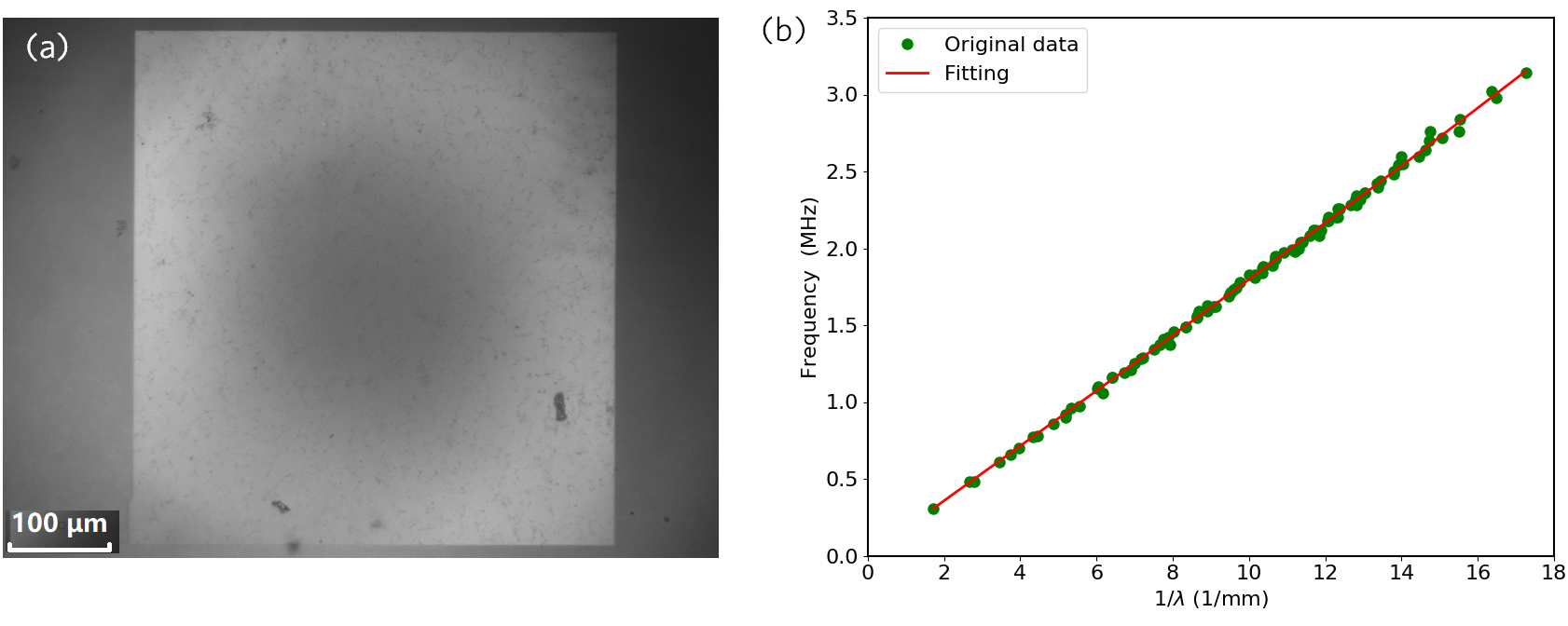}
   \caption[width=0.9\textwidth]{(a) Camera view of the (1,1) mode vibration motion of a G+SiN membrane from the same batch as the sample shown in main text. The motion is captured by stroboscopic light with phase locking during IWLI before the addition of the M leads. (b) By capturing the vibrational motion of different modes, we construct the dispersion relation of the G+SiN membrane before patterning the metal leads. The eigenfrequencies of the different modes (blue dots) are plotted corresponding to their wave number. The dispersion relation is fitted by Eq. \ref{dispersion_relation} and plotted as red solid line.}
   \label{fig:sample3D}
 \end{figure}

As an illustration, Fig.~\ref{fig:sample3D} (a) displays the captured vibrational motion of the ground mode (labeled as (1,1) mode) of a G-covered SiN membrane resonator at a phase $\mathrm{\sim}$ $90^{\circ}$ using stroboscopic light with phase locking of the IWLI. A drumhead vibration deflection can be observed in the center of the membrane. 
\subsection{Mechanical properties: Eigenfrequencies, vibrational patterns, dispersion relation, and nonlinearity}

\noindent The eigenfrequencies of the flexural modes of a rectangular membrane can be calculated using the formula:
\begin{equation}
     \mathrm{\ }f_{m,n}= \sqrt{{({\sigma }_{xx}{m^2}/{L^2_{\mathrm{w}}}+{\sigma }_{yy}{n^2}/{L^2_{\mathrm{h}}})}/{(4\rho )}} \, ,
\label{eq_eigenmodes}
 \end{equation}
here the integers $m$ and $n$ indicate the number of deflection maxima in the two spatial directions of the membrane plane, $L_\mathrm{w}$ and $L_\mathrm{h}$ represent the lateral size of the membrane along the $x$ and $y$ direction, respectively. ${\sigma }_{xx}$ and ${\sigma }_{yy}$ are the residual stress in $x$ and $y$ direction and $\rho$ is the density of the material.\\

\begin{figure}[thp]
  \centering
  \includegraphics[width=\linewidth]{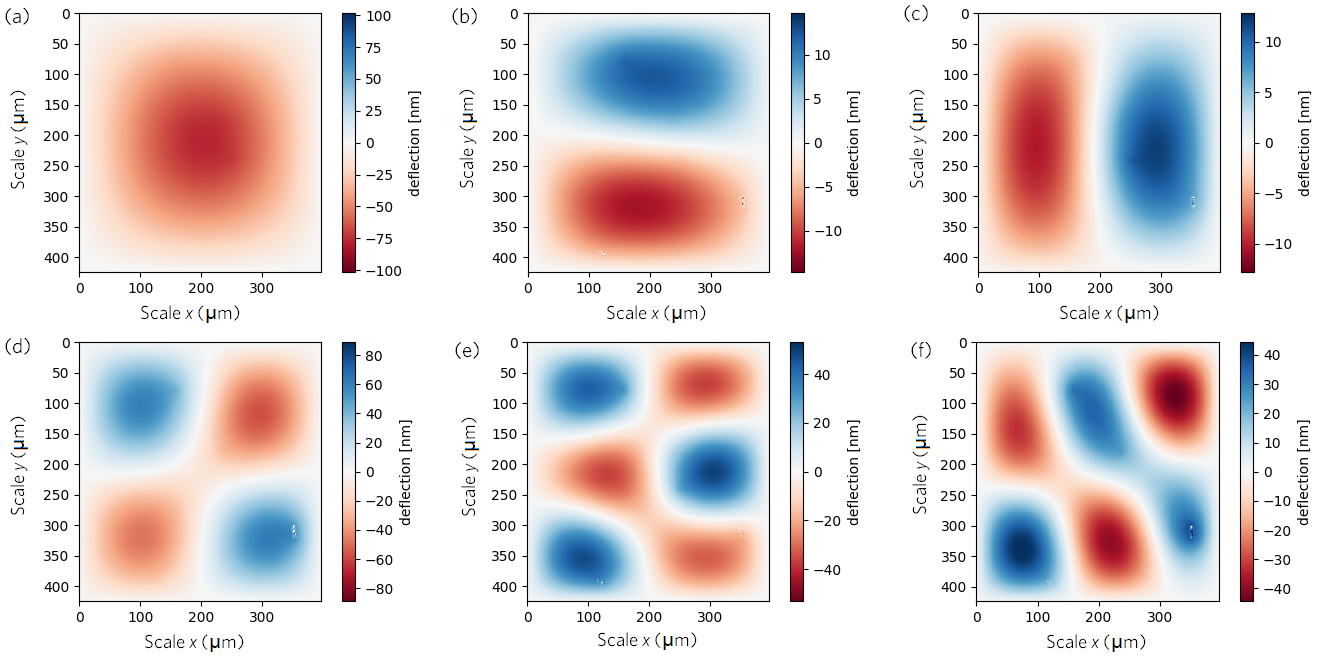}
  \caption{Mode shapes of several representative experimentally captured vibration modes of a G+SiN membrane with slightly smaller lateral dimension.  (a) (1,1) mode at $\sim$ 260\,kHz. (b) (1,2) mode at $\sim$ 460\,kHz. (c) (2,1) mode at $\sim$ 490\,kHz. (d) (2,2) mode at $\sim$ 525\,kHz. (e) (2,3) mode at $\sim$ 760\,kHz. (f) (3,2) mode at $\sim$ 780\,kHz. }\label{fig:G_SiN_modes}
\end{figure}

Figure~\ref{fig:G_SiN_modes} (a) - (f) give examples of measured deflection amplitude profiles, which display the vibration amplitude image of the (1,1) mode (ground mode), (1,2) mode, (2,1) mode, (2,2) mode, (2,3) mode, and (3,2) mode, respectively. 
The dispersion relations are obtained from time series of these images as described in \cite{waitz2012mode}. 
Young's modulus and the residual stress can be determined from the dispersion relation Eq. \eqref{dispersion_relation}:
\begin{equation}
 \omega = \sqrt{\frac{Eh^{2}}{12\rho\left(1-\nu^{2}\right)}k^{2}+\frac{\sigma_{xx}}{\rho}}k.
 \label{dispersion_relation}
\end{equation} 
Here $h$, $\rho$, $\nu$ and $E$ are the thickness, density, Poisson's ratio and Young's modulus of the membrane, respectively.
The stress in the $x$-direction of propagation is $\sigma_{xx}$. The extracted values for the frequency $f$ and the wave number $1/\lambda$ data from $\vec{k}$-space fit method are used to fit the curve. $\omega =2\pi f$ and $k = 2\pi \frac{1}{\lambda}$ are defined. The thickness, density and Poisson's ratio are adapted as $h$ = 110\,nm, $\rho$ = 3180\,$\mathrm{kg/m^{3}}$ and $\nu$ = $0.27$, respectively, in the present case. The estimated Young's modulus and stress are $213$\,GPa and $0.101$\,GPa, respectively.
%
%
%
%
\begin{figure}[thp]
  \centering
  \includegraphics[width=0.9\linewidth]{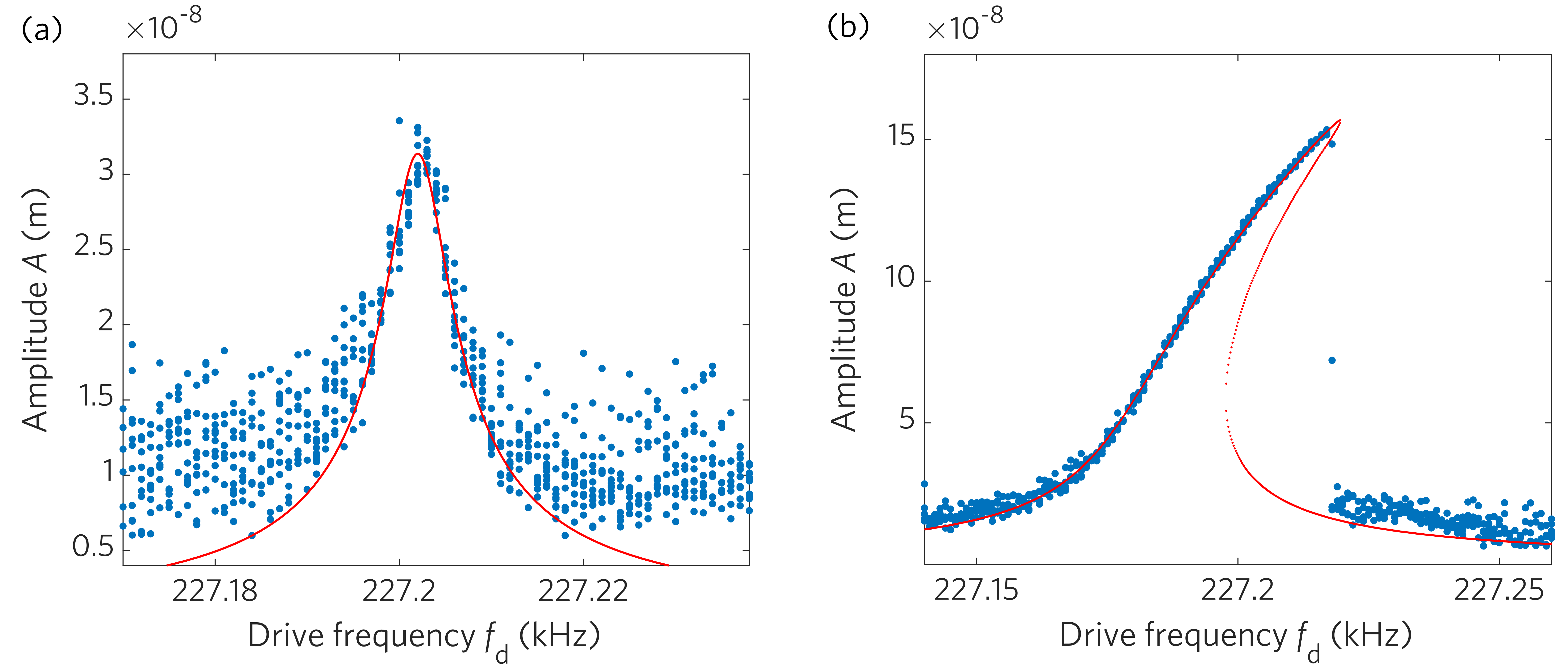}
  \caption{Linear and nonlinear amplitude response of the MGS resonator driven with (a) $V_\textrm{exc}$ = 1.8\,mV and (b) $V_\textrm{exc}$ = 9.1\,mV, respectively. The measured amplitudes (captured by IWLI) are plotted as blue dots, and the theoretical calculation of the Duffing model is plotted as red line. }\label{fig:Duffing_fit}
\end{figure}
%
%
%
The Duffing model is commonly used to describe the nonlinear effects in nanomechanical membrane resonators with the equation of motion:
\begin{equation}
\label{eq:Duff}
\ddot{q}(t)+ 2\pi f_{0}^2 q(t) + 2 \Gamma \dot{q}(t)+ \gamma q^3(t)= F_{\mathrm{d}} \cos \left( 2\pi f_d t \right)
\end{equation}
\noindent where $f_0$,$\Gamma$, $\gamma$, and $F_d$ are the eigenfrequency, damping, nonlinearity and the drive force, respectively. As an example, we analyze the nonlinear amplitude response of the MGS resonator driven with $V_\textrm{exc}$ = 9.1\,mV. The same data as shown in the main text in Fig.~3 (b) is presented in Fig.~\ref{fig:Duffing_fit}. The theoretical modeling indicates that the MGS membrane resonator presents the following mechanical properties: $f_0$ = 227.18\,kHz, $2\Gamma$ = 8.0\,Hz, $\gamma = 3.5 \times 10^{22} ~\textrm{m}^{-2}\textrm{s}^{-2}$, and $F_d$ = 9.8\,N/kg. 

The Duffing nonlinearity of the (1,1) mode can be extracted from the backbone trace characterized by the maximum amplitude and corresponding detuning at a given force, using
\begin{equation}
\label{eq_backbone}
\left(f_{d,max} - f_{0} \right)   = \frac{3\gamma}{8f_{0}} A_{max}^2 \, .
\end{equation}

Typical mechanical response curves of the MGS device from the linear to the nonlinear regime of the (1,1) mode are shown in Fig.~3 (b) in the main text. The extracted backbone trace shown as the red dashed line indicates the nonlinearity $\gamma = 3.64\times10^{22}~\textrm{m}^{-2}\textrm{s}^{-2}$, consistent with the result from the fitting of the Duffing curve.\\

$f_0$ can be calculated by Eq.~\eqref{eq_eigenmodes} with the help of the residual stress determined from the dispersion relation. Moreover, the $\gamma$ can be calculated by the geometric nonlinearity \cite{chen2020determination, liu2018geometric}:
\begin{equation}
     \gamma = \frac{3\pi^4}{16\rho}(\frac{E_{xx}m^4}{L_\textrm{w}^4}+\frac{E_{yy}n^4}{L_\textrm{h}^4}) \,,
\label{eq_nonlinearity}
\end{equation}
\noindent the relation $E_{xx} = E_{yy} = E$  is used considering an isotropic system, and the nonlinearity can be calculated as $\gamma = 3.8 \times 10^{22} ~\textrm{m}^{-2}\textrm{s}^{-2}$, which is similar to the experimental result.\\
\subsection{Sheet resistance and contact resistance of G}
%
%
%
%
\begin{figure}[thp]
  \centering
  \includegraphics[width=0.6\linewidth]{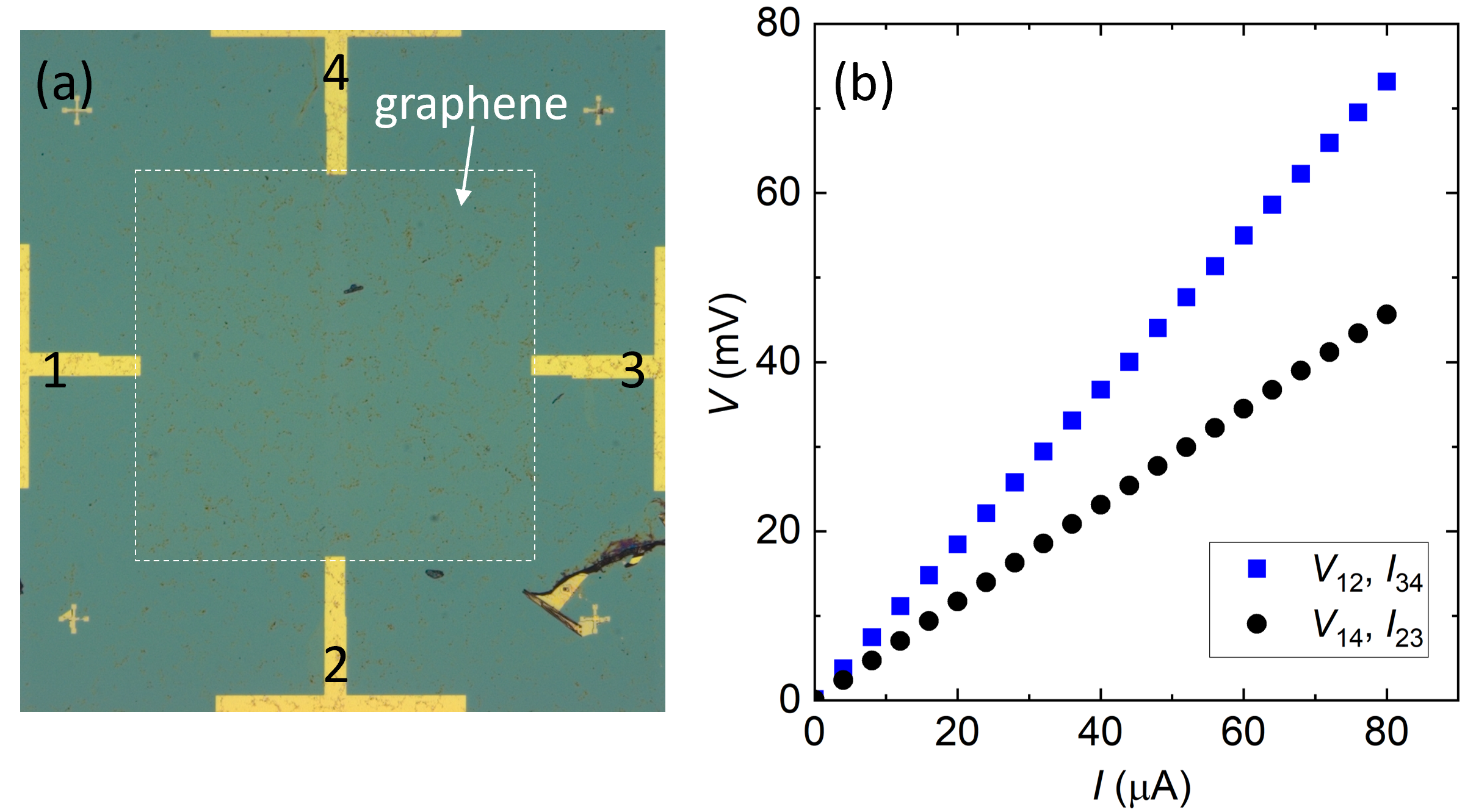}
  \caption{ (a) Optical image of the device used in Van-der-Pauw measurement. (b) $I-V$ measured from two configurations of Van-der-Pauw measurement.  }\label{fig:sheetresistance}
\end{figure}
\begin{figure}[thp]
  \centering
  \includegraphics[width=0.4\linewidth]{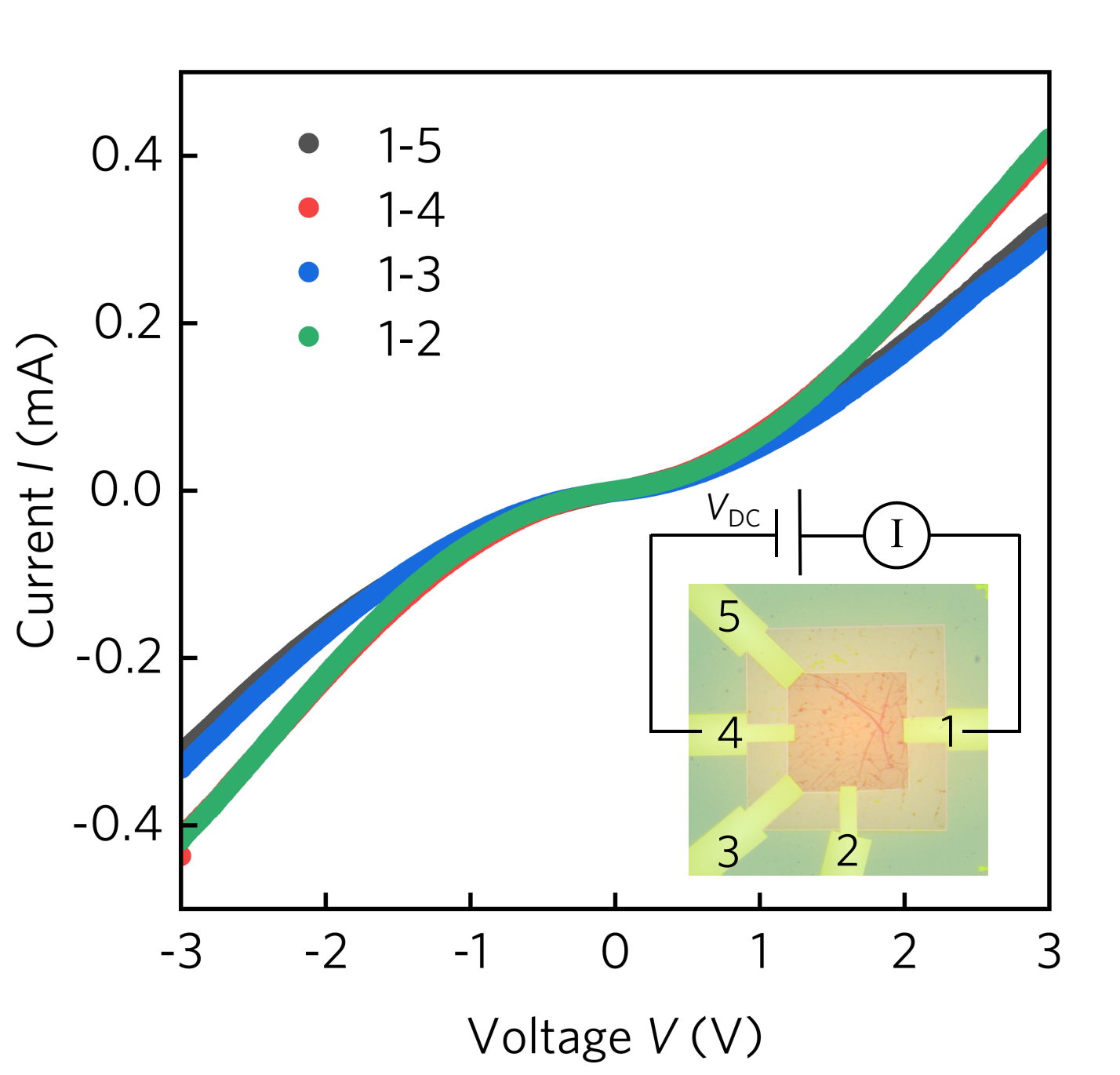}
  \caption{Electric properties of the G. (a) $I-V$ curves between different leads on the membrane surface. Inset: Microscope image of the MGS device with measurement circuit scheme. The metal leads are labeled from 1 to 5; the SiN membrane holds the graphene and metal structures and is clamped by a Si frame.}\label{fig:RU_IV}
\end{figure}

\noindent The total device resistance obtained from the two-point measurement includes the resistance of the G channel and the contact resistances between the M leads and G. To determine their contribution separately, we measured the sheet resistance ($R_{\square}$) of G by the Van-der-Pauw method \cite{ramadan1994van} on another MGS sample, as shown in Fig.~\ref{fig:sheetresistance}. The G layer of the device was grown by the same method, and the fabrication processes of the device are similar with the MGS device investigated in the manuscript. The $R_{\square}$ is calculated to be around 1.65\,k$\Omega$ which is similar to reported values \cite{peng2015sheet}. By using this value and the fact that the G channel is square shaped, we can estimate the channel resistance of G between 1 and 4 to be around 1.65 k$\Omega$. As shown in the two-point measurements between lead 1 to 2, 3, 4, and 5, respectively, in Fig.~\ref{fig:RU_IV}, the two-point resistance of the G-M structure decreases from several tens of k$\Omega$ to around 10 k$\Omega$ as $V_\textrm{DC}$ is increased from 0 to 3\,V, which is much larger than the resistance of the G channel. 
Hence, the contacts between G and M provide the dominating contribution to the total device resistance and thus the heating effects.\\

\begin{figure}[thp]
  \centering
 \includegraphics[width=\linewidth]{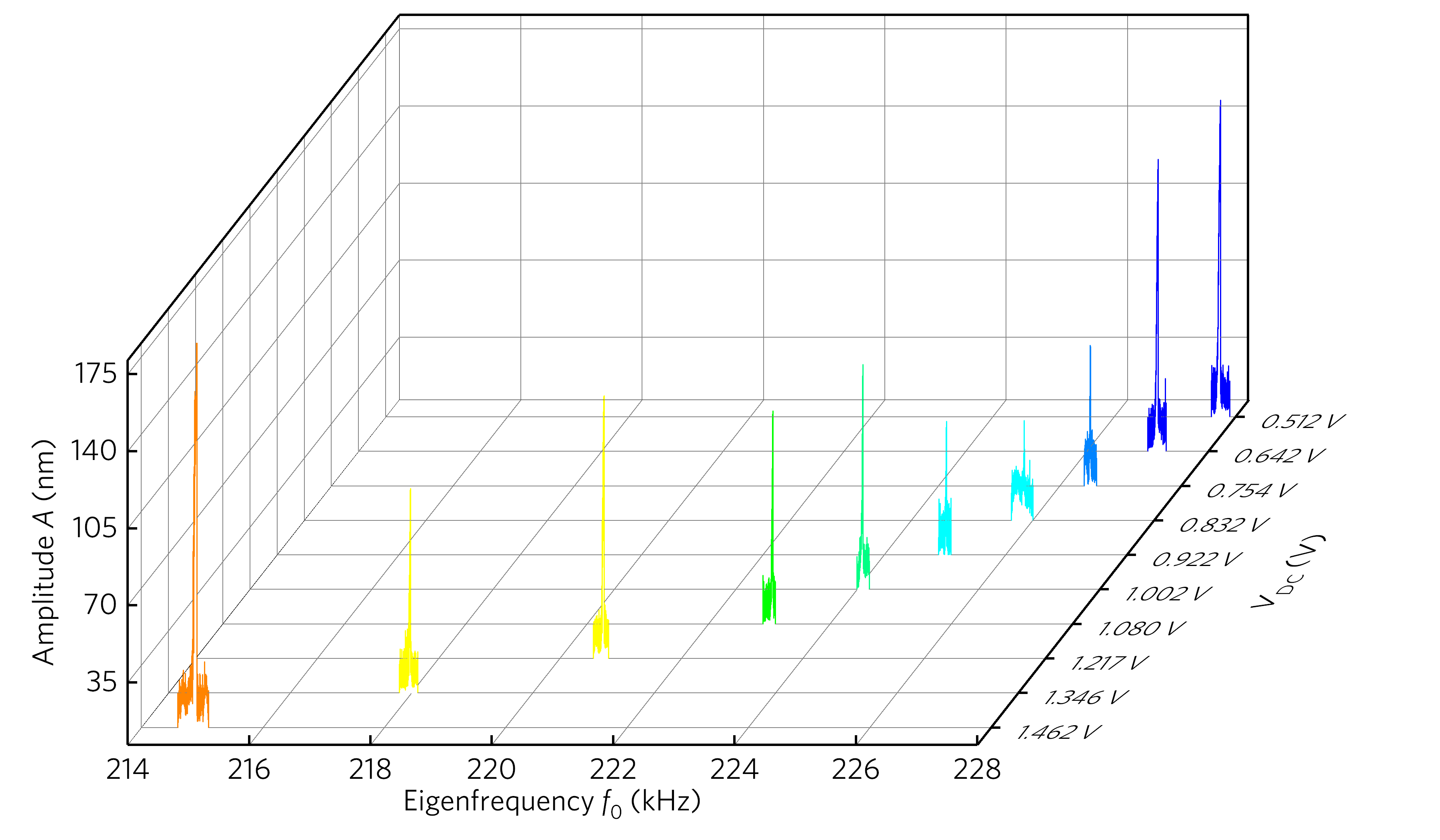}
 \caption{Resonance curves in the Duffing regime measuredby IWLI under different $V_\textrm{DC}$. The observed eigenfrequency shift is roughly proportional to  $P$. }\label{fig:Res_curves}
\end{figure}

In Fig. \ref{fig:Res_curves} we show the resonance curves measured under different $V_\textrm{DC}$ between electrodes 1 and 4 and for excitation voltage $V_\textrm{exc}= 9.1$~mV. By fitting the Duffing model to the experimental data the eigenfrequency was determined and found to shift to smaller values, in agreeement with the data obtained for smaller excitation in the linear regime, shown in Fig. 3 of the main text. 
    
\subsection{Thermomechanically controlled symmetry breaking: static deformation and vibration patterns}
%
%
%
\begin{figure}[thp]
  \centering
  \includegraphics[width=0.9\linewidth]{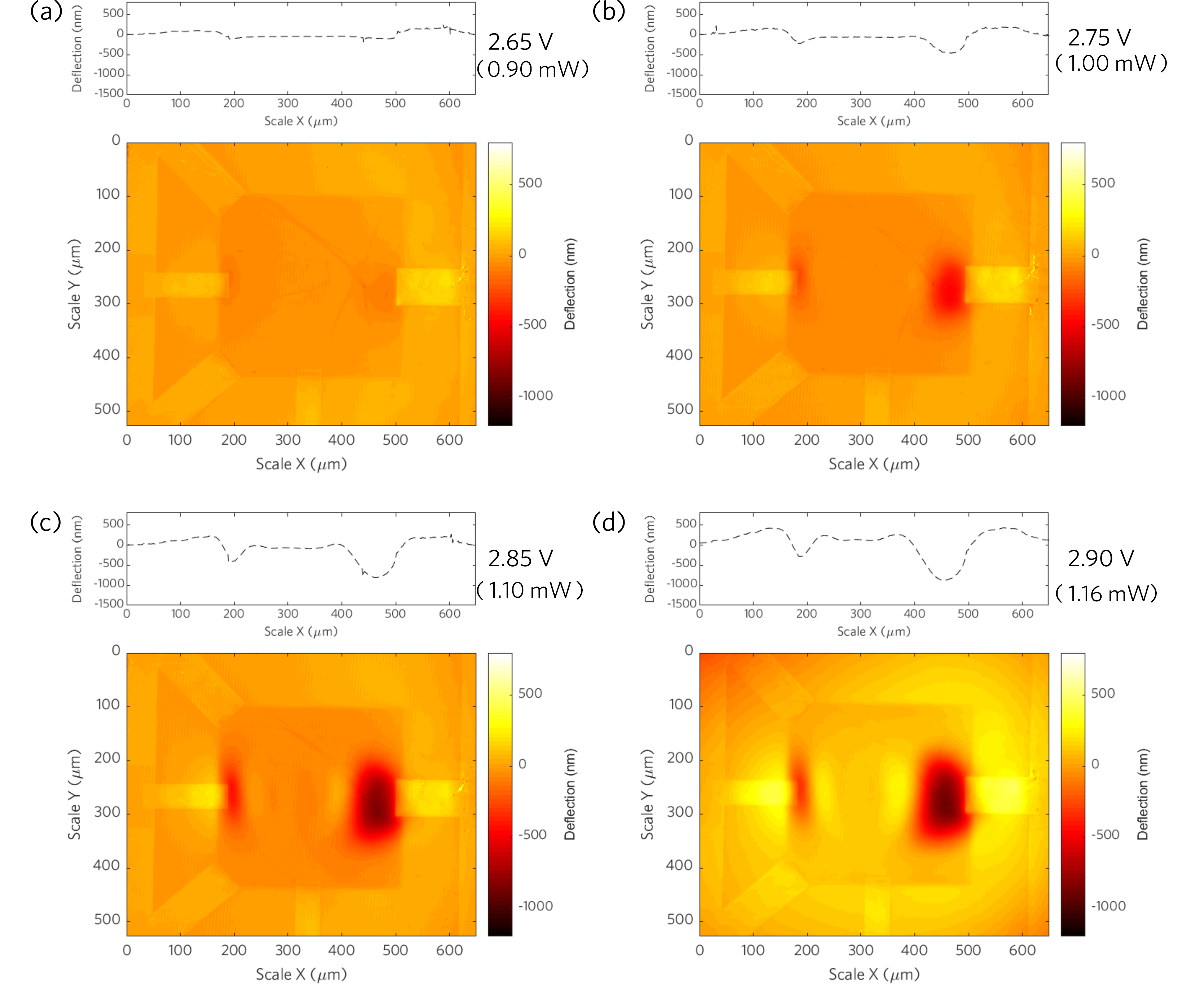}
  \caption{Optical-profilometry characterization of voltage-induced thermal static deformation of the resonator. A line cut of the image at $y = 290~\mu$m is plotted in the upper panel. (a) $V_\textrm{DC}$ = 2.65\,V. (b) $V_\textrm{DC}$ = 2.75\,V. (c) $V_\textrm{DC}$ = 2.85\,V. (d) $V_\textrm{DC}$ = 2.90\,V.}\label{fig:RU-2D}
\end{figure}
%
%
%

\noindent The spatial deflection of the membrane under different $V_\textrm{DC}$ (2.65\,V, 2.75\,V, 2.85\,V and 2.9\,V) applied between leads 1 and 4 is captured by the IWLI and plotted together with line cuts through the center in Fig.~\ref{fig:RU-2D} (a)-(d). The spatial deflection increases with $V_\textrm{DC}$ due to the larger stress generated at the G-M interface. 

The expansion of the M reduces the residual strain along the $x$-axis of the SiN membrane between leads 1 and 4 and also exerts a compressive stress on this area. The compressive stress is also oriented along the $x$-axis from the outer ends (on the frame) of the M leads to their inner ends (on the SiN membrane), squeezing the SiN membrane between leads 1 and 4.
The SiN layer exhibits the most pronounced deflections around the inner ends of the M leads because of the relatively smaller rigidity of the thin SiN layer. When $V_\textrm{DC}$ is larger than 2.75\,V, also the center of the membrane develops corrugations. 
The pronounced deflection is not only observed when $V_\textrm{DC}$ is applied to leads 1 and 4, but also other combinations of the leads show this effect, as shown in Fig.~\ref{fig:RU-eldctrodes}. 

\begin{figure}[htp]
  \centering
  \includegraphics[width=\linewidth]{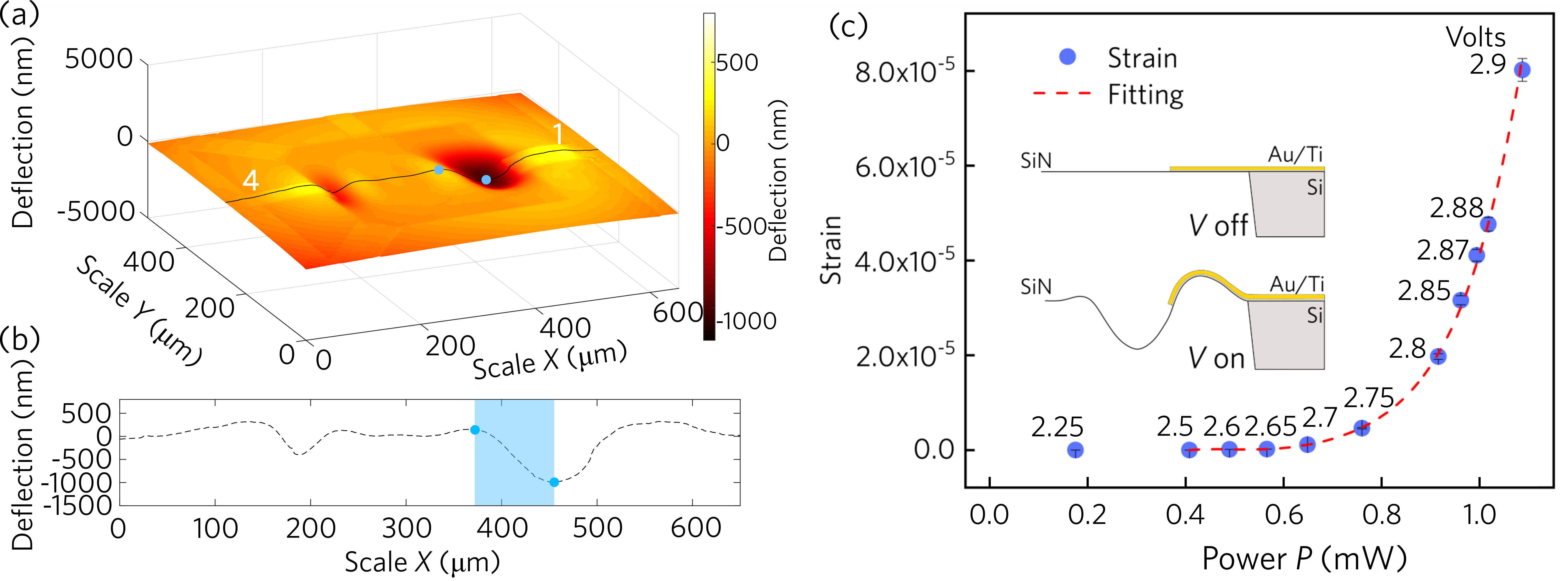}
  \caption{Schemes demonstrating the controlled static deformation and the extracted strain of the SiN membrane.(a) IWLI image of the spatial deflection of the MGS resonator at $V_\textrm{DC}$ = 2.9\,V. (b) Deflection profile of the resonator at the position on the membrane surface indicated by the black line in (a). The selected boundary (370\,$\mu$m, 450\,$\mu$m) of the deflected SiN membrane in $x$-axis for strain calculation is marked as blue dots. (c) The strain is induced by the ETM effect of the MGS structures of lead 1 and 4. The strain is extracted from the spatial deflection along the $x$-axis between 370\,$\mu$m and 450\,$\mu$m. The current is converted from the input $V_\textrm{DC}$ and the measured resistance. The red dashed line indicates a $5^\textrm{th}$ order polynomial fitting.} \label{fig:SM_strain}
\end{figure}

\begin{figure}[thp]
  \centering
  \includegraphics[width=\linewidth]{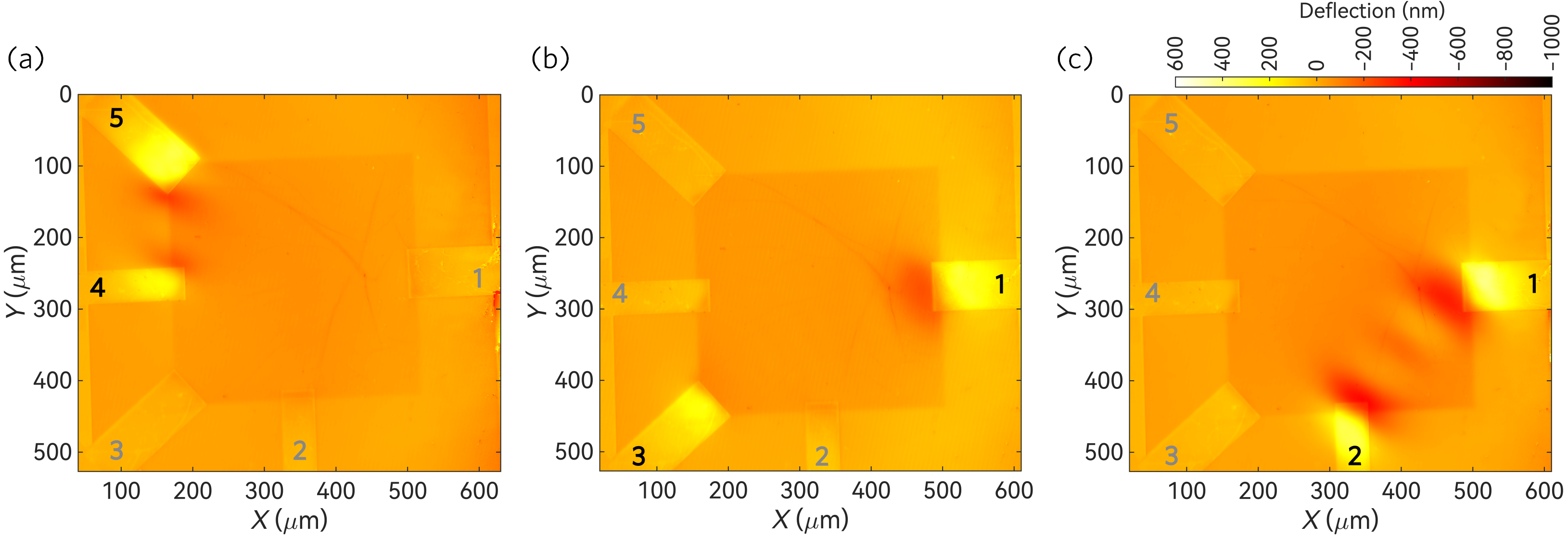}
  \caption{Optical-profilometry characterization of voltage-induced thermal static deformation between different electrodes of the resonator. The labels of the electrodes to which the DC voltage $V_\textrm{DC}$ = 2.75\,V is applied are marked in black, the labels of the unused electrodes are marked in gray. (a) $V_\textrm{DC}$ applied between electrodes 4 and 5. (b) $V_\textrm{DC}$ applied  between electrodes 1 and 3. (c) $V_\textrm{DC}$ applied between electrodes 1 and 2.  The deflections shown in (a) - (c) share the same color scale.}\label{fig:RU-eldctrodes}
\end{figure}

The ETM induced static deformation provides a possible way to monitor the symmetry breaking of the membrane without a capacitive 3D structure (e.g., a back gate located under the membrane) which simplifies the fabrication process. We also extract the strain of the most strongly deformed area and plot it in Fig.~\ref{fig:SM_strain}. The strain is extracted from the spatial deformation along the $x$-axis between of 370\,$\mu$m and 450\,$\mu$m (indicated in Fig.~\ref{fig:SM_strain} (a) and (b)). The strain can be obtained by the following equation: 
\begin{equation}
    \textrm{Strain} = \frac{L_{\textrm{SiN},P}-L_\textrm{SiN}}{L_\textrm{SiN}},
\label{eq_strain}
\end{equation}

\noindent here $L_{\textrm{SiN},P}$ represents the integrated length in the range of the surface between 370 - 450\,$\mu$m under different heating power $P$ generated by $V_\textrm{DC}$, $L_\textrm{SiN}$ represents the original length between 370 - 450\,$\mu$m on the surface (i.e., 80\,$\mu$m) with $P$ = 0. In Fig.~\ref{fig:SM_strain} (c), the extracted strain values with increasing $P$ are plotted as blue dots. A $5^\textrm{th}$ order polynomial function can describe the relation between the ETM-induced strain and the input power, plotted as the red dashed line. 
%
%
%
%
%
%
%
\begin{figure}[thp]
  \centering
  \includegraphics[width=0.8\linewidth]{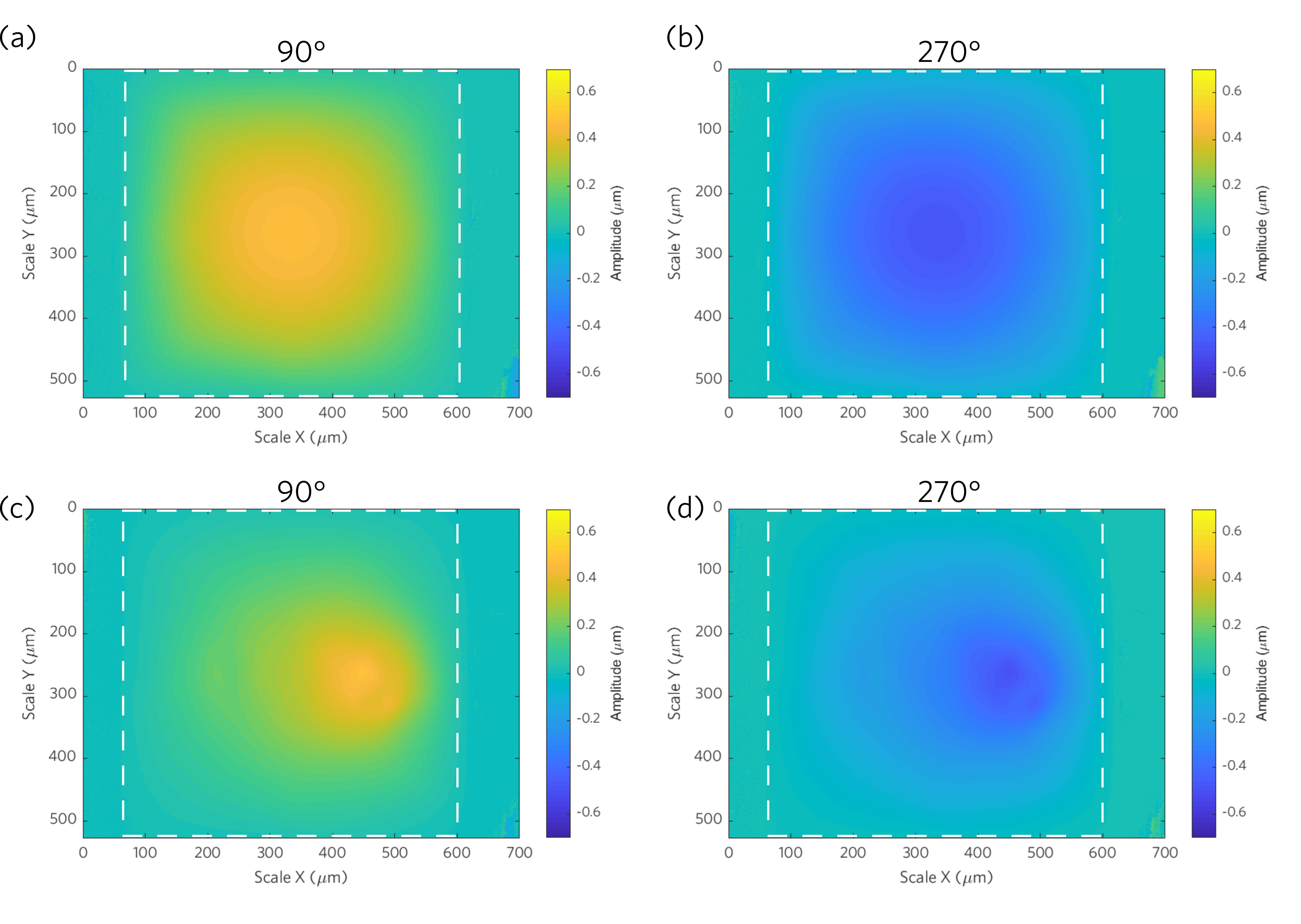}
  \caption{Optical-profilometry characterization of the influence of the symmetry breaking onto the vibrational patterns of the (1,1) mode. (a) IWLI-captured vibration patterns without symmetry breaking ($V_\textrm{DC}$ = 0\,V, phase $\phi = 90^{\circ}$. (b) Without symmetry breaking ($V_\textrm{DC}$ = 0\,V, $\phi = 270^{\circ}$. (c) With symmetry breaking ($V_\textrm{DC}$ = 2.75\,V, $\phi= 90^{\circ}$. (d) With symmetry breaking ($V_\textrm{DC}$ = 2.75\,V, $\phi=270^{\circ}$).} 
  \label{fig:Sym_break}
\end{figure}
%
%

 Here we characterize the spatial deflection of the (1,1) mode with and without symmetry breaking, by stroboscopic IWLI measurements. We applied $V_\textrm{DC}$ = 0\,V and $V_\textrm{DC}$ = 2.75\,V to leads 1 and 4 to demonstrate the ETM controlled symmetry breaking. The device is driven by the piezo with $V_\textrm{exc} = 0.5$\,V. The drive frequency is swept up from below to slightly above the respective eigenfrequency (the eigenfrequency is shifted to smaller values due to the heating effects as well as by the symmetry breaking introduced by $V_\textrm{DC}$). When $V_\textrm{DC}$ = 0\,V, there is no overall heating and symmetry breaking in the system. The spatial deflection of the vibrational motion of the (1,1) mode shows a usual sinusoidal envelope of the amplitude distribution, which is symmetric with respect to the $x$ and $y$ axis. The two patterns are captured at the phase of $\phi=90^{\circ}$ and 270$^{\circ}$ by stroboscopic measurement, shown in Fig.~\ref{fig:Sym_break} (a) and (b), respectively.
An overall heating and symmetry breaking is observed by applying $V_\textrm{DC} = 2.75$\,V 
to the leads 1 and 4. As shown in Fig.~\ref{fig:RU-2D} (b), the membrane shows a static deformation with maximal height $\approx 500$\,nm  (representing symmetry breaking). A voltage $V_\textrm{exc}$  of the same size
is applied to the piezo and the frequency is swept around $f_0$ which is significantly tuned by the overall heating and symmetry breaking. The patterns are captured at $\phi= 90^{\circ}$ and 270$^{\circ}$ by stroboscopic measurement as well and are shown in Fig.~\ref{fig:Sym_break} (c) and (d). Significant symmetry breaking in the dynamic regime can be observed. The maximum deflection amplitude is no longer in the center of the membrane but shifted to the side of lead 1, where the most pronounced static deformation was obtained by $V_\textrm{DC}$.
\subsection{Model for curvature radius: modified Timoshenko model}
\noindent We built our model based on certain simplifications motivated by the geometry and properties of our experimental device:\\
\noindent (1)	The bending of the MGS structure is subject to a uniform heating from $T_0$ to $T$;\\
\noindent (2)	The metal lead is considered as one continuous layer and the expansion effect of G is neglected because of its ultrasmall thickness;\\
\noindent (3)	The MGS structure is considered as a cantilever with prestress (equal to the residual stress of the SiN membrane) applied on the free end of SiN;\\
\noindent (4)	The difference in the thermal expansion coefficients remains constant during heating such that the friction at the supports can be neglected.\\

\indent If the thermal expansion coefficient of the metal layer and SiN, $\alpha_\textrm{M}$ and $\alpha_\textrm{SiN}$, are different, the heating will produce bending of the MGS structure similar to the Timoshenko bilayer cantilever \cite{timoshenko1925analysis}. $E_\textrm{M}$ and $E_\textrm{SiN}$ denote their Young's modulus, $h_\textrm{M}$ and $h_\textrm{SiN}$ their thicknesses, $h = h_\textrm{M} + h_\textrm{SiN}$ the total thickness. The width of the cantilever is taken as equal to unity.

The following analysis is made on the assumption that cross sections along the $x$-axis of the M leads originally were planar. The cross section perpendicular to the $x$-axis remains constant during the bending, i.e. the curvature in the $y$-axis is negligible due to the small width of the M leads.

Since $\alpha_\textrm{M} > \alpha_\textrm{SiN}$,  the M leads elongate more than SiN and therefore the deflection is convex upwards \cite{timoshenko1925analysis}. For the metal on the convex side, all forces acting on the cross section can be represented by an axial compressing force $P_\textrm{M}$ and bending moment $M_\textrm{M}$. The forces acting over the section of SiN on the concave side can be represented by an axial tensile force $P_\textrm{SiN}$, a residual stress $P_\textrm{res}$ = $h_\textrm{SiN}\sigma_\textrm{SiN}$, and bending moment $M_\textrm{SiN}$. Since it is assumed that no other external forces except residual stress are acting on the cantilever, all forces acting over any cross section of the cantilever must compensate each other. Hence we obtain:
\begin{equation}
\label{eq:Timo_1}
P_\textrm{M} = P_\textrm{SiN} + P_\textrm{res} = P
\end{equation}
\noindent and
\begin{equation}
\label{eq:Timo_2}
\frac{P_h}{2} = M_\textrm{M} + M_\textrm{SiN} .
\end{equation}

\noindent Letting $r_c$ represent the curvature radius of the cantilever, $E_\textrm{M}I_\textrm{M}$ is the flexural rigidity of the metal and $E_\textrm{SiN}I_\textrm{SiN}$ represents the flexural rigidity of the SiN. We define $M_\textrm{M}=\frac{E_\textrm{M}I_\textrm{M}}{r_c}$ and $M_\textrm{SiN}=\frac{E_\textrm{SiN}I_\textrm{SiN}}{r_c}$, combining with Eq. \eqref{eq:Timo_2} we get: 
\begin{equation}
\label{eq:Timo_3}
\frac{P_h}{2}=\frac{E_\textrm{M}I_\textrm{M}+E_\textrm{SiN}I_\textrm{SiN}}{r_c}. 
\end{equation}
And by using Eq. \eqref{eq:Timo_1} and Eq. \eqref{eq:Timo_3},  $P_\textrm{M}$ and $P_\textrm{SiN}$ adopt the following form:
\begin{equation}
\label{eq:Timo_4}
P_\textrm{M}=\frac{2(E_\textrm{M}I_\textrm{M}+E_\textrm{SiN}I_\textrm{SiN})}{r_c h} ,
\end{equation}
\begin{equation}
\label{eq:Timo_5}
P_\textrm{SiN}=\frac{2(E_\textrm{M}I_\textrm{M}+E_\textrm{SiN}I_\textrm{SiN})}{r_c h}+P_\mathrm{res} .
\end{equation}

Another equation for calculating $P$ and $r_c$ can be obtained from the consideration of deformation. On the surface of M and SiN, the unit elongation occurring in the longitudinal direction must keep equal. This results in the equation
\begin{equation}
\label{eq:Timo_6}
\alpha_\textrm{M}\left(T-T_0\right)-\frac{P_\textrm{M}}{E_\textrm{M}h_\textrm{M}}-\frac{h_\textrm{M}}{2r_c}=\alpha_\textrm{SiN}\left(T-T_0\right)+\frac{P_\textrm{SiN}}{E_\textrm{SiN}h_\textrm{SiN}}+\frac{h_\textrm{SiN}}{2r_c}
\end{equation}
Using Eq. \eqref{eq:Timo_4} and Eq. \eqref{eq:Timo_5},  Eq. \eqref{eq:Timo_6} becomes
\begin{equation}
\label{eq:Timo_7}
\left(\alpha_\textrm{M}-\alpha_\textrm{SiN}\right)\left(T-T_0\right)-\frac{P_\mathrm{res}}{E_\textrm{SiN}h_\textrm{SiN}}=\frac{2\left(E_\textrm{M}I_\textrm{M}+E_\textrm{SiN}I_\textrm{SiN}\right)}{r_c h} \left(\frac{1}{E_\textrm{M}h_\textrm{M}}+\frac{1}{E_\textrm{SiN}h_\textrm{SiN}}\right)+\frac{{(h}_\textrm{M}+h_\textrm{SiN})}{2r_c} ,
\end{equation}

\noindent which leads to

\begin{equation}
\label{eq:Timo_8}
\frac{1}{r_c}=\frac{\left(\alpha_\textrm{M}-\alpha_\textrm{SiN}\right)\left(T-T_0\right)-\frac{P_\mathrm{res}}{E_\textrm{SiN}h_\textrm{SiN}}}{\frac{2\left(E_\textrm{M}I_\textrm{M}+E_\textrm{SiN}I_\textrm{SiN}\right)}{h} \left(\frac{1}{E_\textrm{M}h_\textrm{M}}+\frac{1}{E_\textrm{SiN}h_\textrm{SiN}}\right)+\frac{{(h}_\textrm{M}+h_\textrm{SiN})}{2}} .
\end{equation}
With the definitions $\frac{h_\textrm{M}}{h_\textrm{SiN}}=h'$, $\frac{E_\textrm{M}}{E_\textrm{SiN}}=E'$, here $I_\textrm{M}=\frac{{h_\textrm{M}}^3}{12}$ and $I_\textrm{SiN}=\frac{{h_\textrm{SiN}}^3}{12}$, the curvature of the MGS structure follows the general equation:
\begin{equation}
\label{eq:Timo_9}
\frac{1}{r_c}=\frac{(6\left(\alpha_\textrm{M}-\alpha_\textrm{SiN}\right)\left(T-T_0\right)-\frac{\sigma_\textrm{SiN}}{E_\textrm{SiN}}){(1+h')}^2}{h(3\left(1+h'\right)^2+\left(1+h'E'\right)\left(h'^2+\frac{1}{h'E'}\right))} .
\end{equation}
Here the temperature of the MGS structure at different $V_\textrm{DC}$ is estimated by applying Fourier's law. To calculate the temperature of lead 1, the following parameters are used: $T_0$ = 26\,\textcelsius, $L_\textrm{M}$ = 110\,$\mu$m, $W_\textrm{M}$ = 60\,$\mu$m, $h_\textrm{M}$ = 30\,nm, and $\kappa$ = 318\,Wm$^{-1}$K$^{-1}$ (for Au). Considering the resistance contribution of the contact resistances, thermal radiation and heat conductance of the SiN layer perpendicular to the MGS, the heat exerted to the MGS structure is estimated to be roughly 85$\%$ of the total generated heat. The eigenfrequency shift $\delta f_0$ $\simeq$ 40\,kHz can be estimated from Fig.~3 (a) in the main text with $V_\textrm{DC}$ = 2.25\,V, typically corresponding to a temperature shift around 80\,K ($\sim$ 500\,Hz/K \cite{yang2023quantitative}) and the actual temperature averaged over the surface of the device is $\sim$ 105 \textcelsius, by taking the heat dissipation into account, similar to the estimation (85 \textcelsius) by Fourier's law. \\

%
%
\subsection{COMSOL simulation of temperature distribution and deformation of the MGS}

\noindent To testify the rationality of our simplifications in the analytic model and support our experimental results better, the Finite-Elements-Analysis (FEA) software package COMSOL Multi physics (version 6.1, academic license from Eidgenössische Technische Hochschule Zürich (ETH Zürich) is utilized to simulate the temperature distribution and deformation caused by the ETM effects on the MGS. 
The dimensions of the device under test are presented in Table~\ref{dimensions}, which matches the structure and dimensions of the MGS device shown in Fig.~\ref{fig:RU_IWLI} and Fig.~1 in the main text. Since the quantum transport properties of a monolayer G cannot adequately be described by COMSOL, and in the present experiment we just use it as heater with a particular resistance, we mimicked it by a thin layer with thickness 1\,nm  with a resistivity adjusted such that it results in a resistance close to the experimental value.
 A prestress of 0.101\,GPa (as measured in the experiment) is applied to the four edges of the SiN membrane. Details of the physics modules setup and corresponding domains can be found in Table~\ref{physics}.\\
%
\begin{figure*}
\includegraphics[width=0.6\linewidth]{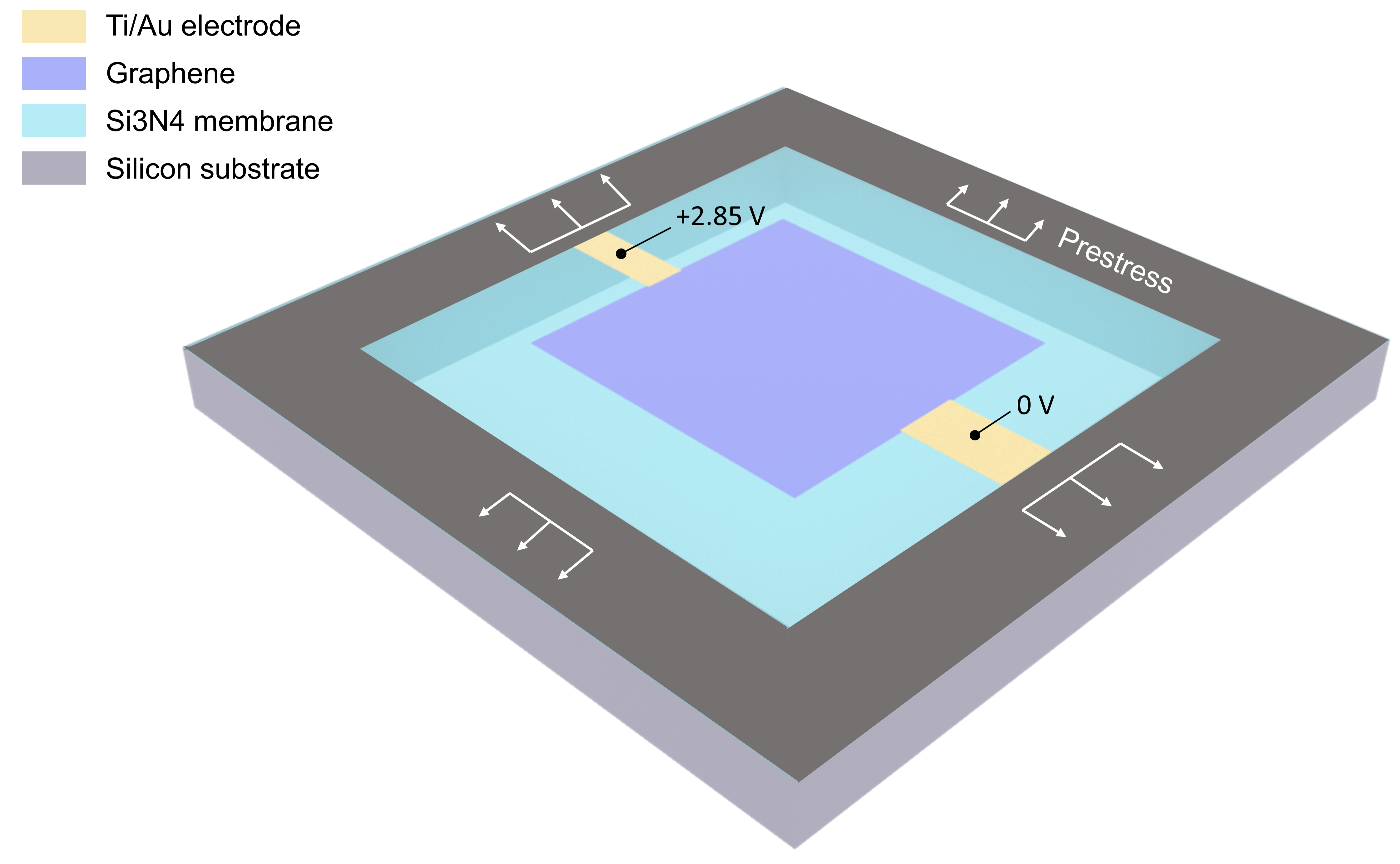}
\caption{\label{fig:setup}Device characteristics and simulation setup}
\end{figure*}

\begin{figure*}
\includegraphics[width=0.8\linewidth]{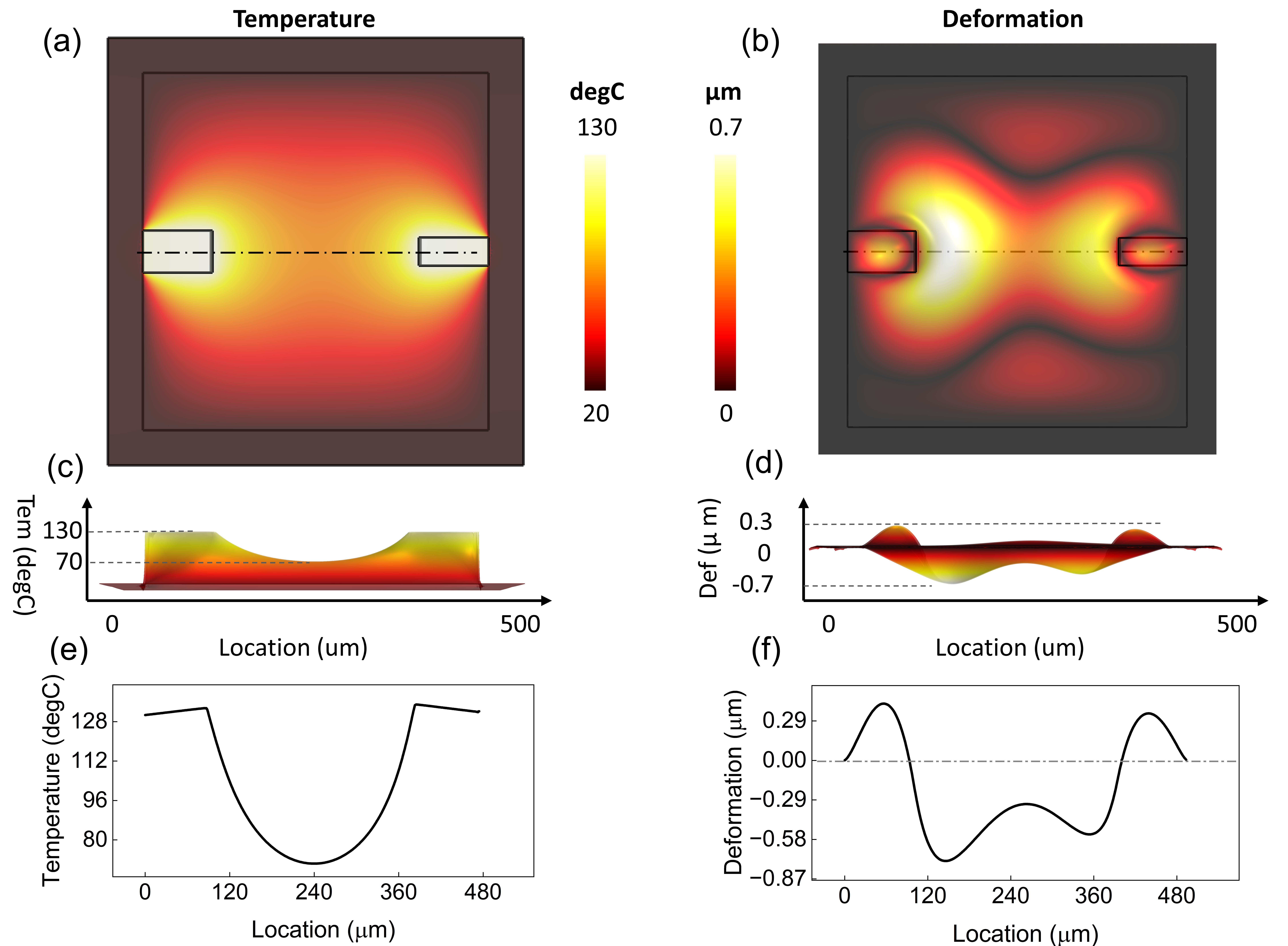}
\caption{\label{fig:res}The temperature distribution and deformation of the SiN membrane. Panels (a) and (c) depict the temperature distribution of the membrane from a top view and side view, respectively. Panel (e) displays the temperature profile along the path between the two electrodes. The deformation of the membrane is represented in panels (b) and (d) from a top view and side view, respectively. Finally, panel (f) exhibits the deformation profile along the path between the two electrodes.}
\end{figure*}

A static voltage $V_\textrm{DC}$ = 2.85\,V is applied to the M electrodes. In order to approximate the relatively large contact resistance between G and M observed in our experiments, we enhanced the resistivity of both electrodes to 3.6×$10^{-8}$ $\Omega$m.
 With these parameters we achieved a generated current of approximately 0.315 mA, which is close to the experimental value (around 0.34 mA).
 \begin{table}
 \caption{Sample properties \label{dimensions}}
 \begin{ruledtabular}
 \begin{tabular}{ccc}
 \textbf{Domain}&\textbf{Dimensions ($\mu m$)}&\textbf{Thermal expansion coefficient}\\
 Graphene&$300\times300\times0.001$& $-3.2\cdot 10^{-6} / K$\\
 Electrode 1&$100\times60\times0.025$ (Ti) + $100\times60\times0.025$ (Au)&Au ($14.2\cdot 10^{-6} / K$)\\
 Electrode 2&$100\times40\times0.025$ (Ti) + $100\times40\times0.025$ (Au)&Ti ($7.6\cdot 10^{-6} / K$)\\
 SiN membrane&$495\times512\times0.11$&$3.2\cdot 10^{-6} / K$\\
 \end{tabular}
 \end{ruledtabular}
 \end{table}
In Fig.~\ref{fig:res}, we present the temperature distribution and deformation of the SiN membrane. The highest temperature is concentrated on both electrodes, reaching approximately 130\,\textcelsius, while the lowest temperature of about 70\,\textcelsius is found in the middle of the membrane. In terms of deformation, the two electrodes deflect upwards, while the membrane close to the electrode area deflects downwards.
The spatial distribution of temperature and deformation agrees well with our experimental results, providing strong support for our simplified theoretical analysis outlined in the main text.
 \begin{table}
 \caption{Physics module setup\label{physics}}
 \begin{ruledtabular}
 \begin{tabular}{cc}
 \textbf{Physics module}&\textbf{Domains}\\
 Solid Mechanics&Graphene, Ti/Au electrode, SiN membrane, Si substrate\\
 Electrical Currents&Graphene, Graphene, Ti/Au electrode, SiN membrane\\
Heat transfer in Solids&Graphene, Ti/Au electrode, SiN membrane, Si substrate 
 \end{tabular}
 \end{ruledtabular}
 \end{table}
%
%
%
%
%
\newpage
%
\subsection{Destructive test: conductance and static deformation}
%
%
\begin{figure}[thp]
  \centering
  \includegraphics[width=0.8\linewidth]{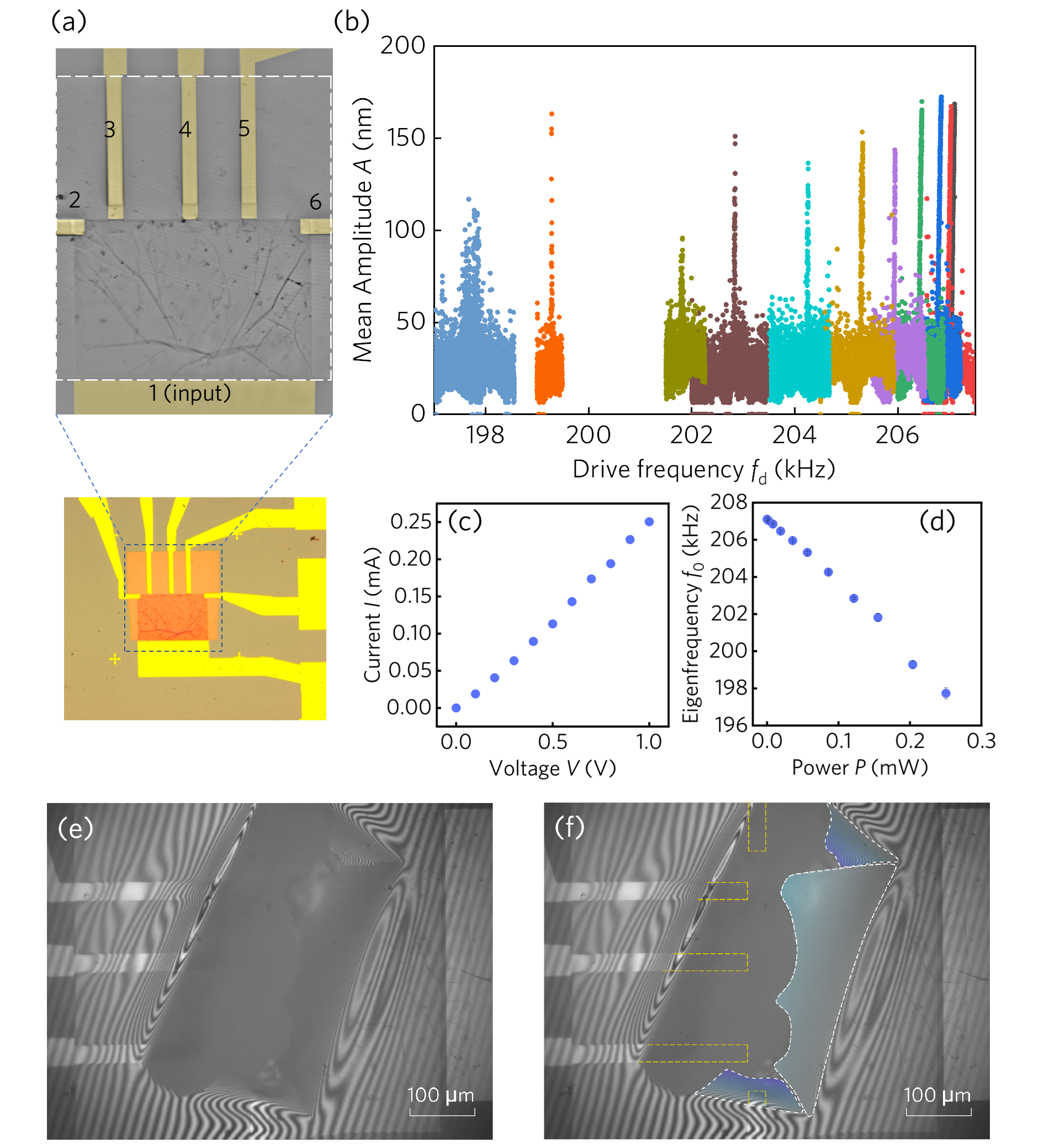}
  \caption{DC voltage tuning of the eigenfrequency of the (1,1) mode of the resonator. (a) Light microscope image of the sample and a zoom of the sample (framed by the blue dashed line) captured by IWLI. The SiN membrane area is framed by the white dashed line. (b) IWLI measured amplitude response curves of the (1,1) mode under different input $V_\textrm{DC}$. (c) The two-point $I-V$ between lead 1 and all others used together as counter electrode. (d) The relation of $I^2$ (proportional to the input power) and $f_0$.  (e) Static interference picture of the sample captured by IWLI after the application of large $V_\textrm{DC}$ to lead 1 w.r.t. all others on ground potential and continuously increasing it up to 7.0\,V, until the membrane was destroyed. Figure rotated counterclockwise by 90° w.r.t. (a). (f) Same image as in (e) with markups of the boundary of the broken membrane (white dashed line). Areas of different pieces of the broken membrane are colored to indicate the most strained area. The yellow dashed lines mark the position of the electrical leads. }\label{fig:SM_RD_f0}
\end{figure}
%
%
\noindent We here demonstrate a destructive test on a similar sample, a G-covered SiN prestressed membrane, with 6 electrodes on the surface of the SiN membrane, as shown in Fig.~\ref{fig:SM_RD_f0} (a), captured by IWLI. The only difference to the sample used in the main text is the geometry of the patterned G and M leads. The white dashed line indicates the boundary of the SiN membrane; the rectangle area with wrinkles corresponds to the patterned G and the yellow colored structures indicate the M leads, labeled from 1 to 6. For a better view, the optical image of the sample under the microscope is shown in Fig.~\ref{fig:SM_RD_f0} (c). 
The $V_\textrm{DC}$ is applied between the electrode 1 and the others (2,3,4,5, and 6 on the same ground potential). A linear eigenfrequency shift can be observed while $V_\textrm{DC}$ increases from 0.0 V up to 1.0 V in steps of 0.1 V. The resonance curves are captured by IWLI at different drive frequency as shown in (b) and the measured current and the fitted $f_0$ for corresponding $V_\textrm{DC}$ are plotted in (d) and (e) as blue dots. The linear dependence is fitted to the $I-V$ curve and $I^2$ - $f_0$ relation, as blue dashed line plotted in Fig.~\ref{fig:SM_RD_f0} (d) and (e). The $I-V$ and the corresponding $f_0$ shifts due to the heating show similar behavior compared to the sample shown in the main text.
However, when we increase $V_\textrm{DC}$ up to around 6\,V, the leads 2, 3, 4, 5 and 6 deform tremendously, the leads drag the membrane down and finally break the membrane when $V_\textrm{DC}$ reaches 7 V. The captured image after the membrane breakdown is shown in Fig.~\ref{fig:SM_RD_f0} (e). In panel (f) we painted the broken boundary with different transparent colors for highlighting the shape.   
%
%
%
\subsection{On-surface actuation of a MGS resonator }
\begin{figure}[thp]
  \centering
  \includegraphics[width=0.5\linewidth]{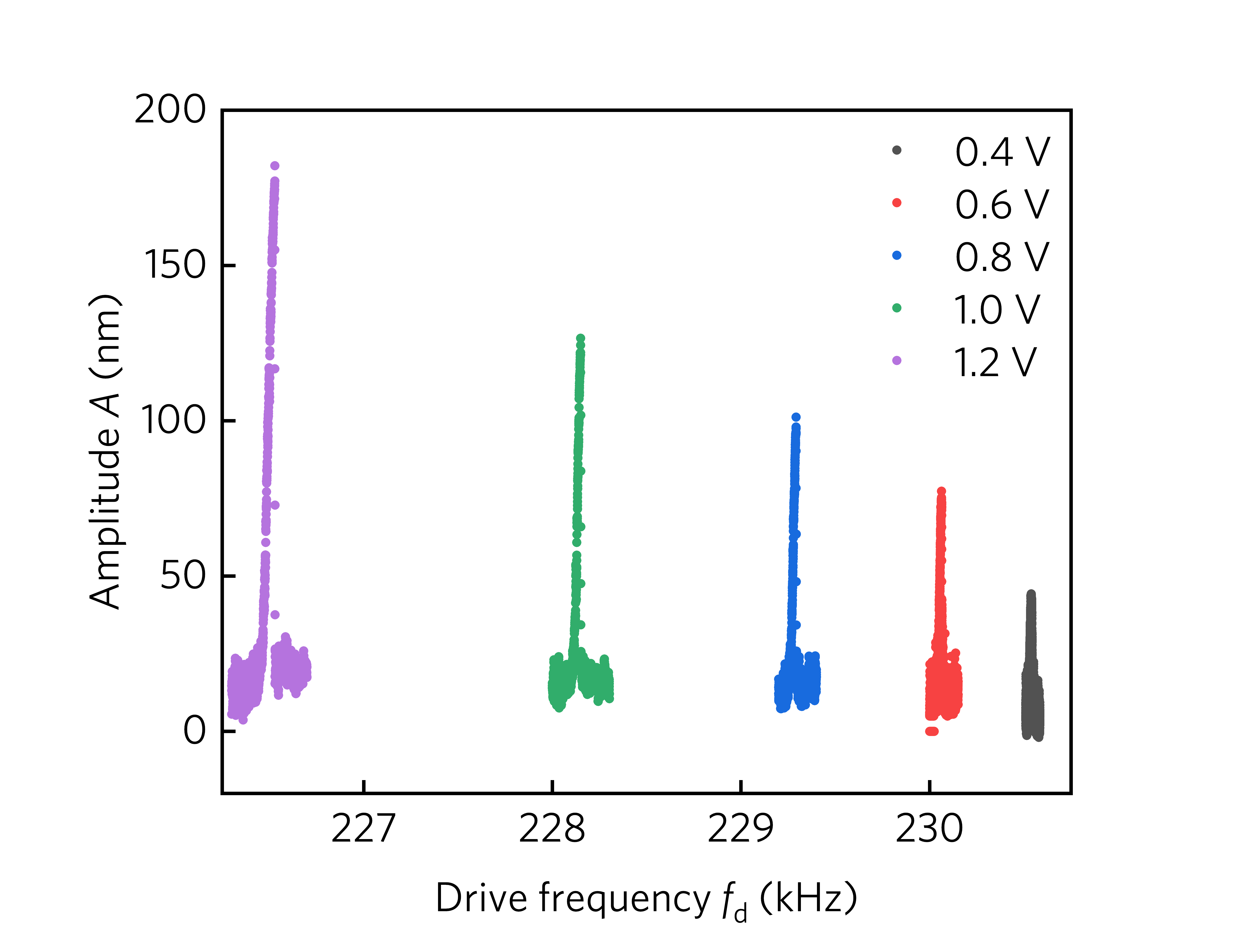}
  \caption{Measured vibrational response curves of the (1,1) mode as function of the drive frequency. Same data as presented as function of the detuning in Fig.~4 in the main text.}
  \label{fig:ABS_AC_drive}
\end{figure}

\noindent Different $V_\textrm{AC}$ are applied between leads 1 and 4. We sweep the frequency of $V_\textrm{AC}$ close to the eigenfrequency of the (1,1) mode, hence we excite the membrane into vibration. We here show the same data as shown in Fig.~4 in the main text, but now plotted as function of the drive frequency. With increasing $V_\textrm{AC}$ the effective power dissipated between leads 1 and 4 produces an overall heating effect which leads to a red shift of $f_0$. The maximum amplitude response of the driven mode increases while increasing the intensity of $V_\textrm{AC}$.
%
%
%
%
\bibliographystyle{apsrev4-1}
\bibliography{Dispersion_si}